\documentclass[a4]{article}
\usepackage{graphicx}
\usepackage{subcaption}
\usepackage{algorithm}
\usepackage{algorithmic}
\usepackage[utf8]{inputenc}
\usepackage{hyperref}
\usepackage{xcolor}
\usepackage[official]{eurosym}

\title{The Reverse File System: Towards open cost-effective secure WORM storage devices for logging}

\author{Gorka Guardiola Múzquiz  \\
\and
Juan González-Gómez \\
\and
Enrique Soriano-Salvador
\\
~\\
EIF, Universidad Rey Juan Carlos
~\\
{\small \texttt{\{gorka.guardiola,juan.gonzalez.gomez,enrique.soriano\}@urjc.es}}
}

\begin{document}
\maketitle

\begin{abstract}
Write Once Read Many (WORM) properties for storage devices are desirable to
ensure data immutability for applications such as secure logging,
regulatory compliance, archival storage, and other types of backup
systems.  WORM devices guarantee that data, once written, cannot be
altered or deleted. However, implementing secure and compatible WORM
storage remains a challenge.  Traditional solutions often
rely on specialized hardware, which is either costly, closed,
or inaccessible to the general public.  Distributed approaches,
while promising, introduce additional risks such as denial-of-service
vulnerabilities and operational complexity.  We introduce Socarrat, a
novel, cost-effective, and local WORM storage solution that leverages
a simple external USB device—specifically, a single-board computer
running Linux with USB On-The-Go (OTG) support. The resulting device
can be connected via USB, appearing as an ordinary external
disk formatted with an ext4 or exFAT file system,
without requiring any specialized software or drivers.  By isolating the
WORM enforcement mechanism in a dedicated USB hardware module,
Socarrat significantly reduces the attack surface and ensures that even
privileged attackers cannot modify or erase stored data.  In addition
to the WORM capacity, the system is designed to be tamper-evident,
becoming resilient against advanced attacks.
This work describes a novel approach,
the \emph{Reverse File System}, based on inferring the file system
operations occurring at higher layers in the host computer where Socarrat
is mounted.
The paper also describes the current Socarrat prototype,
implemented in Go and available as free/libre software. Finally, it provides
a complete evaluation of the logging performance on different
single-board computers.
\end{abstract}

\section{Introduction}

Write Once Read Many (WORM) storage devices are a key component
in systems that require immutable data retention. Once data is written
to a WORM device, it cannot be modified or deleted, though it remains
accessible for unlimited number of read operations.
The WORM property is essential for
a wide range of applications, including secure logging.
Data regulations (see for example~\cite{GDPRregulation,SECregulation})
increasingly require the assurance of long-term data retention,
accountability, integrity, and verifiable data migration.
WORM devices play a critical role in fulfilling these regulatory
and operational requirements.

Despite the apparent simplicity of the WORM model, implementing compatible,
secure, and reliable WORM storage is a complex issue. Traditional approaches
depend on specialized hardware. These solutions are
often expensive, inaccessible to the general public, or difficult to
integrate with modern computing environments.
Distributed systems offer
alternative approaches but introduce their own challenges, including
increased complexity and susceptibility to denial-of-service attacks.
Moreover, they are not suitable for disconnected or intermittently
connected systems.
In this work, we focus on local solutions for logging purposes.

A critical threat arises when an attacker gains control
over the host machine\footnote{From now on, we will refer to the machine to
which the storage-providing device is connected as \emph{the host}.}.
In such scenarios, conventional file system
protections (e.g. the \emph{append-only} and \emph{immutable}
attributes of the standard Linux file system)
are insufficient: If the attacker elevates privileges, she
can reconfigure the file system, modify or delete files, tamper with
data blocks, or even reformat the storage device. These issues
highlight the need for a WORM solution that is resilient to the complete
compromise of the host and does not rely solely on the file system enforcement
mechanisms (attributes, permissions, etc.).

To address these challenges, we propose Socarrat, a novel and
cost-effective WORM storage solution implemented as a standalone USB
device. Socarrat is built on a single-board computer (e.g. a Raspberry Pi)
running Linux with USB On-The-Go (OTG) support~\cite{utg}.
This device is identified as a mass storage device that can be mounted
on any conventional operating system (Windows, Linux or MacOSX)
like any ordinary USB external disk formatted with a standard file system.
At present, our system supports two different file systems: ext4~\cite{ext4}
(the standard file system for Linux systems) and exFAT~\cite{exFAT}.
This way, for approximately 100 \euro{}\footnote{The cost of a Raspberry Pi 4,
an SD Card, and an external 1 TB USB disk. No costs are incurred from software
licensing, as our system is based on free/libre technology.}, Socarrat provides a
practical and secure WORM solution accessible to a broad range of users.

At the core of Socarrat is a novel architectural approach we call
\emph{the Reverse File System}. It
infers a superset of the high-level file system operations (those performed by the
file system driver executing in the host).
These operations are reconstructed by analyzing  the low-level disk blocks
written to the storage device (i.e. our
single-board computer) through the USB link.
This component focuses exclusively on append-only write operations of any size
(from 1 byte to the maximum write limit) to
designated log files (those with WORM capabilities),
discarding all other modifications. It ensures
that only valid, sequential log entries are retained, while unauthorized
writes (i.e. modifications to the previously written data)
are neutralized\footnote{Note that the idea
of a \emph{Reverse File System} is general and not limited
to our application (i.e. building WORM devices).
The approach is powerful and can be applied to
other kind of applications (debugging, fuzzing, etc.).}.

Building such a system presents significant technical challenges.
The research problem arises from the fact that a single partition
cannot be mounted concurrently by two operating systems
(the host and the Socarrat device), due to
obvious synchronization and consistency issues.
Our system exports a raw block volume to the host,
and must accurately interpret the blocks it receives
to determine whether they correspond to valid append operations
on protected log files. This requires a deep understanding of the file
system internals, including the structures (e.g. inodes),
data blocks (e.g. extents), and journaling mechanisms.
The system must also detect and
handle partial writes and coalesce multiple block updates into coherent
write operations.

In addition, Socarrat provides forward integrity to the data already
written to the secured logs.
It uses our previous system, SealFS~\cite{sealfsv1,sealfsv2},
to enable tamper-evident, authenticated logs.
If the attacker is able to circumvent the protections and attack
our storage device over the USB link or by other means,
any modification to committed data in the logs will still be detected
by a verification tool.

\subsection{Example scenario and operational procedure \label{scenario}}

To illustrate the use of our system, consider the following scenario:
Alice connects a USB black box (i.e. the Socarrat device) to her server (i.e. the host).
This device has been initialized previously by the Auditor (Alice or a third party).
The operating system of the host detects a USB mass
storage device formatted as an ext4 file system. Then, this
drive is mounted in the system. At the mount point, there is a file
named \texttt{log} (set up at initialization stage).
The applications running in the server are able to
use this volume as a regular one, by performing the traditional system
calls to work with files.
The only difference is that, when the applications
access the \texttt{log} file, only read operations and append-only
write operations are effective; the rest of the write operations over the
\texttt{log} file (and its metadata) are discarded.
Later, the Auditor can extract the \texttt{log}
file from the black box, along with an additional file that authenticates
all the entries added to the log for this period. Using a diagnostic
tool, the Auditor  can verify that the log data
corresponds to the entries made during the operational period, ensuring
that no records have been deleted or tampered with.

Therefore, there are three distinct stages:

\begin{enumerate}
	\item Configuration: Configure which log files will be provided and
	generate the necessary data to initialize the tamper-evident mechanisms.
	This is done by the Auditor.

	\item Logging: Alice connects the device and it works as an already formatted
	mass storage device.

	\item Auditing: The Auditor
	extracts the logs from the device and verifies their integrity.
\end{enumerate}

\subsection{Contributions}

The contributions of this paper are:

\begin{enumerate}
	\item The introduction of a novel approach, the \emph{Reverse File System},
	to implement storage devices based on USB OTG.
	Such devices can be used as ordinary external
	disks, but permit us to control the operation performed
	over the data objects provided at higher levels in the host machine.

	\item The design of an architecture, Socarrat, that
	enables cost-efficient tamper-evident WORM devices by integrating
	the Reverse File System approach and SealFS, as an out-of-the-box
	feature for any modern operating system.

	\item The implementation details of our current research prototype,
	which supports ext4 and exFAT file systems.

	\item The evaluation of this prototype running on different
	types of single-board computers, including the analysis
	of the results for the experiments conducted to measure
	its performance for logging purposes.
\end{enumerate}

\subsection{Organization}

The rest of the manuscript is organized as follows:
Section~\ref{related} discusses the related work;
Section~\ref{model} briefly describes our threat model;
Section~\ref{arch} explains the general architecture of the system;
Section~\ref{response} describes the indicators of compromise and possible responses;
Section~\ref{implementation} provides the implementation details of the current prototype;
Section~\ref{evaluation} describes the experiments we conducted and the evaluation results;
and
Section~\ref{conclusions} presents the conclusions.

\section{Related work \label{related}}

\subsection{Hardware}

Old WORM systems used continuous feed printers~\cite{Bellare97forwardintegrity},
tapes, and optical devices like the DEC RV20 laser drives with 2 GB RV02K-01 disks
(1988), Sony WDD disks (1992) and Sony WDA Writable Disk Auto Changer jukebox (1992).
Tape and optical media still exist, but it is now largely confined to
niche or legacy applications.
For example, Sony launched
their third-generation optical disks (5 TB per unit) in 2020.
These products were discontinued in 2024.
IBM and Hewlett Packard also produce tapes for WORM storage (e.g.
IBM LTO WORM Data Cartridge and HPE StoreEver LTO‑8 Ultrium 30750).
The price of the LTO‑8 Ultrium 30750 tape drive is $\approx$ 4000 \euro{}.
These devices have been falling
into disuse for decades and are not as readily available as standard drives.
We present a solution based on conventional storage
devices and low-cost single-board computers: We are able to
build a WORM device for $\approx$ 100 \euro{}.

Some lesser-known manufacturers offer WORM disks.
For example, Greentec\footnote{https://greentec-usa.com/products} offers a
product called WORMdisk ZT Storage,
compatible with file systems such as NTFS, exFAT, and ext4.
Another example are the Flexxon’s WORM SD
Cards\footnote{https://www.flexxon.com/write-once-read-many-worm-memory-card/}.
The underlying technology of these devices is closed and obtaining
further details about the architecture and pricing appears to be
challenging.
Our motivation is to provide an open, low-cost, accessible, and
flexible solution with a clear delineated set of guarantees.

There is significant ongoing research into new materials  (e.g. organic components)
for manufacturing new types of WORM memories, but no general-purpose
products based on these technologies are
commercially available yet (see for example ~\cite{leppaniemi2012roll,HSU2024174252}).
That corpus of work lies outside the scope of this study.

\subsection{Drivers and file systems}

Ordinary file systems like Linux's ext4, include mechanisms
to force files to be append-only and immutable. Nevertheless, if the machine is
compromised and the attacker has administrator privileges, those attributes
can be disabled. We aim to provide a hardware device to ensure those properties
(and provide extra guarantees, like forward integrity).

Finlayson et al. presented Clio~\cite{clio} in 1987. Clio is
an extension for the V-System file system, specifically designed
for logging.
It used a combination of optical drives and magnetic disks to
provide append-only files.

In 1991, Quinlan developed a WORM system for the Plan~9 operating
system~\cite{plan9worm}. It was also based on an optical disk jukebox and a
magnetic disk. The magnetic disk, which is used as a cache,
provides the files for common use (read and write).
The optical devices are used to provide read-only
snapshots of the file system.
Later, other file systems focused on snapshots emerged.
For example, the WOWSnap file system~\cite{5617032} uses two partitions, one for ordinary files and other
for WORM files to provide snapshots.
Ext3cow~\cite{ext3cow} was another open-source file versioning and snapshot
system, derived from ext3.
Those two systems differ in the approach. WOWSnap implements a ROW
(Redirect On Write) policy, while Ext3cow implements a COW (Copy On Write)
policy.
These systems do not provide append-only files.

Blutopia~\cite{blutopia}, a system implemented by an author of this manuscript
in 2007,
provided immutable file systems through stackable block device. It requires
periodic snapshots and it is not suitable for append-only files.

Wang et al.~\cite{1410755} proposed a preliminary approach to build
append-only storage system with conventional disks.
In this work, they categorize the WORM systems in three groups: physical
WORM (e.g. optical drives), coded WORM (e.g. operating system drivers)
and software WORM (e.g. applications). They proposed to modify (i) the operating
system block device drivers for the disks to force append-only write operations
and (ii) the disk's firmware.
While (i) represents a realistic approach, (ii) poses a significant
challenge to achieve, given that the disks' firmware is entirely
closed source.
In this work, the authors explain their approach
and provide some pseudocode,
but they do not describe a functional prototype. The approach for (i) is
to manage two partitions, one for data and the other one for metadata that
maps the regions already written (to detect rewrites).
Although the authors state that the approach is independent of the type
of file system, it is not clear how the file system
metadata  and internal structures
(i.e. inodes, superblock, etc.) could be handled.
We provide a complete and functional prototype: We have built our own \emph{``disk''} with a
single-board computer, to control its \emph{``firmware''} (i.e. Socarrat),
which is based on our \emph{Reverse File System} idea.
This way, (i) is not necessary.
Debiez et al.~\cite{wormpatent} patented a similar method, based on modifications of
the device driver (i). Note that it is not secure in case of total
compromise of the host system,
because the attacker could replace the device drivers.

As far as we know, the concept of a \emph{Reverse File System} is novel,
and there are few comparable systems.
The two most similar examples we have identified are closely related,
with one having been directly inspired by the other.
Vvfat~\cite{vvfat} is a block driver served by
Qemu. It serves a virtual VFAT file system in which the
files represent the files of an underlying directory and follow
them when read an written.
Inspired on this, there is the \texttt{nbdkit} floppy plugin~\cite{nfp}.
First of all, those systems are not used to provide WORM devices or secure logs.
Second, they  implement a kind of \emph{synthetic reverse file system}.
They are unconstrained by a liveness property: They
only need to guarantee that the underlying files are completely synchronized
when they are unmounted. In contrast,
we must continuously extract valid operations to
function properly.
Last, our
\emph{Reverse File System} watches the data structures present in a
block device image, while those two \emph{synthetic reverse file systems}
make up the data structures on demand.

\subsection{Distributed approaches}

There are numerous distributed approaches to provide logging systems,
leveraging cloud computing (see for example ~\cite{loggly,stackdriver}).
In some cases, these systems offload the problem to another machine over
the network. This just shuffles the problem elsewhere:
Servers that store the logs can be attacked as well.
In addition, the networked system architecture significantly
increases the attack surface compared to our approach, which is
confined to the USB mass storage interface.
Furthermore, distributed systems may add extra latency, especially in the
presence of lengthy consensus or proof of work algorithms.
We aim to provide a local solution that can be used
by systems operating offline by basing it on a
small USB device.


There are some cloud services that offer distributed
WORM archival capabilities.
For example, AWS S3 Object Lock~\cite{s3} provides WORM properties
with immutable files, but it does not support appendable files.
IBM Cloud Object Storage also provides
non-erasable and non-rewritable data objects.
Files stored in such an immutable can be set to
immutable or append-only~\cite{ibmobject}.

Samba, the open source implementation of the SMB/CIFS network file system,
provides WORM functionality on the client side
with a module named Samba vfs\_worm~\cite{sambaworm}.
This module is an additional layer that forces WORM
semantics. In case the host is compromised, the module can be disabled.
Moreover, on the server side, the files are mutable.

NetApp SnapLock~\cite{snaplock} is a commercial solution that
offers WORM storage for network file systems, such as NFS
and SMB. It supports WORM append-only files.
It also stores cryptographic hashes to ensure the integrity of the immutable
data in the storage servers.
In this system, the data is appended in chunks of 256 KB.
We provide append-only files without this limitation (the
append-only write size is not fixed).
In any case, as we stated before, we aim to provide a local solution without
network dependencies.

Huawei also offers network storage devices (the OceanStor series)
that include a system named HyperLock.
This system can store regular files, append-only files, and immutable
files (what they call WORM files)~\cite{Huawei}.

GlusterFS~\cite{davies2013scale,7600187} is a storage system
that exports a network file system from
multiple existing volumes (e.g. ext4), hosted on different servers.
It enables the creation of different types of compound volumes
(distributed, replicated, dispersed, etc.). In this system
it is possible to create a WORM volume that provides append-only
files~\cite{glusterfsdocs}.
The Gluster native client is only available for Linux. Other
systems can access the file system through NFS and SMB.

Quinlan et al. created Fossil~\cite{quinlan2003fossil}
and Venti~\cite{quinlan2002venti}. Fossil is a file system that
can create archival snapshots,
which are backed up to a Venti file server~\cite{quinlan2002venti}.
Venti is a content addressed block storage system that uses a Merkle tree of
SHA1 hashes to index read-only blocks.
Only snapshots are read-only. Although Fossil provides append-only
files (through an attribute), if the corresponding blocks are not
stored in Venti yet (i.e. they are not part of a snapshot), the data can be
modified. While Fossil and Venti were created as part of a distributed
system, Plan~9~\cite{pike1990plan}, they can certainly be configured locally in
the same machine. The problem is that, in that case, when there is a total
compromise of the system, the attacker can rewrite the Venti blocks, recreating
the Merkle tree of hashes and the WORM property is compromised.
In addition, Venti handles complete blocks, like other systems discussed before.

Other distributed file systems, like the Hadoop distributed File
System (HDFS)~\cite{hdfs}
and Google FS (GFS)~\cite{gfs}, are also based on immutability.
These file system have no concept of a change to a complete file.
When a file is created, is append-only~\cite{10.1145/2857274.2884038}.
Those systems are specifically designed to store
large data objects for big data applications, its use just for storing
secure logs may be overkill.

\subsection{Forward integrity}

There are multiple distributed systems that provide forward integrity
based on blockchain or distributed ledger
technologies (see for example~\cite{blackboxlog2019,10.1007/978-3-030-16187-3_9,medusa,logsentinel}).
Again, those solutions are not suitable for disconnected or loosely connected
scenarios.
We achieve forward integrity locally, using a previous system called
SealFS~\cite{sealfsv1,sealfsv2}, that combines ratcheting~\cite{Bellare97forwardintegrity}
with storage based tamper-evident logging.
For a more in-depth study on tamper-evident logs, please refer
to~\cite{sealfsv1,sealfsv2}.

\section{Threat model \label{model}}

The asset to be secured consists of the logs generated by
the processes running on the host system.
To formalize the guarantees we aim to provide, we introduce the
Continuous Printer Model (CPM). This model, inspired by the old approach
to provide WORM logs (i.e. using a continuous printer to generate a hard copy of the logs),
captures the essential properties of secure WORM storage in adversarial
environments:

\begin{itemize}
	\item Forward Immutability: Once data is committed to disk,
	it cannot be deleted or overwritten. There is no guarantee that:
		\begin{itemize}
			\item Spurious or bad data will not be committed in the future
			\item The system will not be stopped from \emph{printing} (i.e. crashed).
		\end{itemize}

  	In any case, Immutability of already committed data is always preserved:
	What is \emph{printed} cannot be \emph{unprinted}.

	\item Liveness: Data is committed frequently enough to ensure timely persistence.
\end{itemize}

Additionally, with the SealFS component, Socarrat also provides:

\begin{itemize}
	\item Forward Integrity: Even if the host or the Socarrat device
        are compromised, previously committed data remains
	verifiable.
\end{itemize}

Note that what we have called \emph{Forward Immutability} is a storage property (data cannot be deleted or rewritten), whereas,
Forward Integrity is a cryptographic property (data cannot be deleted or modified without detection).

\subsection{Threats}

We consider two distinct types of attacks:

\begin{itemize}

        \item Logical attacks targeting the host.
	Upon exploitation, the attacker
	may control the host machine and execute code at any privilege
	level. Thus, she can write any information to the logs, change the
	metadata, modify the file system configuration,
	unmount the volumes, try to format it, write directly to
	the block devices, etc.

 	\item Physical or logical attacks to the Socarrat device.
  	Upon the attack, the adversary has access to the internal Socarrat system,
	its private local storage devices, etc.
\end{itemize}

\subsection{Dependencies}

The system has a single requirement: The Socarrat device must have a
storage devices with enough capacity to store the logs,
the authentication data, and
the cryptographic keys used by SealFS,
over extended operational periods.

The hardware platform for the Socarrat devices is standard (e.g.
an ARM-based single-board computer equipped with conventional
SSD, SD or NVME storage components).

\subsection{Assumptions}

\begin{itemize}
        \item The host may operate either online or offline.
	Both hardware and software components are assumed to function
	correctly and remain trustworthy until compromised by an
	adversary through system exploitation, process corruption,
	or the execution of malicious code.

	\item The Socarrat device is connected to the host
	through a USB cable. The integrity of the USB link is preserved
	and it only provides the mass storage device within specifications.
	Both hardware and software components are assumed to function
	correctly and remain trustworthy until compromised by an
	adversary through system exploitation, process corruption,
	or the execution of malicious code.

	\item The attacks are carried out during Stage 2, that is,
	the normal operation of the system (the \emph{logging}
	stage detailed in Section~\ref{scenario}).

	\item The Socarrat device employs a secure deletion procedure, ensuring that
	once data is removed, it becomes irretrievable from
	all storage layers, including persistent memory, cache, and disk.

	\item The Auditor performs log verification on an independent,
	reliable, and trusted system connected to the Socarrat device.
\end{itemize}

\subsection{Mitigation}

Against remote or local logical attacks targeting the host,
the first line of defense of the Socarrat device is the USB link.
Therefore, the attack surface is limited solely to the USB interface.
The block device interface,
which we extend to the other machine via the USB mass storage device, acts
as a choke point: The set of operations on it is limited.

Internally, the \emph{Reverse File System} acts as a kind of data diode, by only
honoring certain operations (i.e. writes at the end of the file)
and ignoring the rest.
The information finally stored in authenticated logs flows in only one
direction (toward the logs themselves).
If the USB link does not fall, we provide the CPM guarantees.

Upon attack detection, Socarrat can apply different policies (they will
be discussed in Section~\ref{response}).

Against logical attacks targeting the Socarrat device
system (e.g. the attacker is able to exploit vulnerabilities of the
USB link in order to execute malicious code in the Socarrat device),
or in case of physical attacks to tamper with its storage components,
the system just provides forward-integrity. If the data already stored
in the log files is manipulated, the Auditor will be able to notice it
and disregard the logs (i.e. the committed logging data
cannot be deleted or counterfeit without being detected).

\section{Architecture \label{arch}}

\begin{figure}[th]
\begin{center}
	\includegraphics[width=\columnwidth]{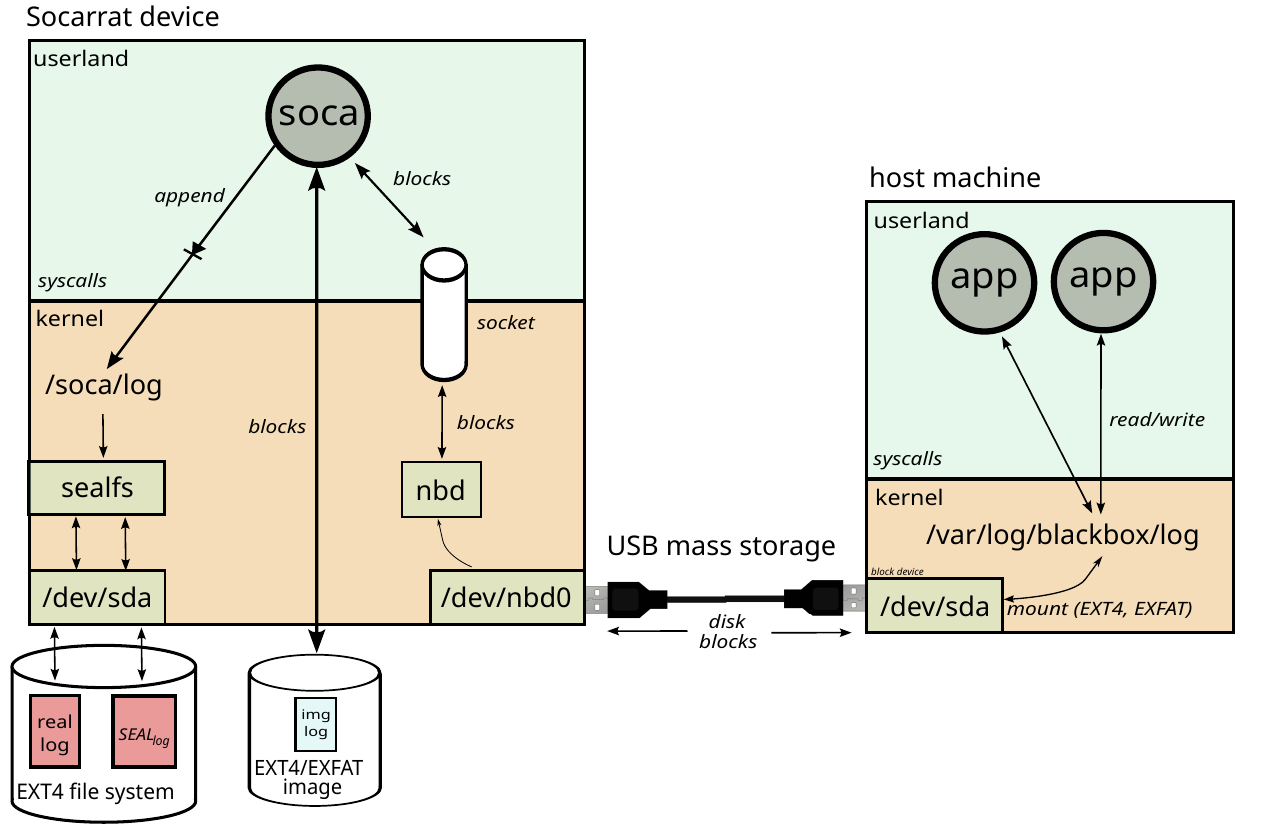}
	\caption{General architecture of the Socarrat system. \label{diagram}}
\end{center}
\end{figure}

An overview of the architecture of the system can be seen in Figure~\ref{diagram}.
The host is an ordinary computer or another kind of system (e.g. a robot)
running Linux (or any other operating system that supports ext4 or exFAT
file systems), which is executing standard
user-space applications. From now on, we will suppose that the selected
file system is ext4 for the sake of simplicity.

The host has mounted the file system provided by the USB mass storage device,
represented by an ordinary block device (\texttt{/dev/sda}
in this example). In the figure, we suppose that the file
system is mounted in \texttt{/var/log/blackbox}. This file system provides
some preconfigured log files (\texttt{log} in this example).
The ext4 driver of the host will read and write blocks out of
\texttt{/dev/sda}.

From the point of view of the user, the Socarrat device
is an ordinary USB disk formatted as ext4.
Those applications use the files normally. In the case of the
logs, they use them as ordinary append-only files. These applications
do not need to use any special library or framework to access the files.

In the Socarrat device, there is a user-space program named \texttt{soca}
that implements the \emph{Reverse File System}.
\texttt{Soca} serves the blocks of the Socarrat device. It processes the read
and write operations over blocks, and infers the operations that are being
performed by the ext4 driver of the host.
Note that it does not infer the exact original file system operations  performed by the applications in the host (e.g.
a write system call on a file), but a subset of the operations limited by the
information available (i.e. the blocks it receives), possibly aggregated.
It interacts with two storage elements provided by the local operating system:

	\begin{itemize}
		\item An ext4 image file or a disk partition
		that contains the blocks that it
		serves. From now on, we will assume that we are using an image
		file to simplify the explanations\footnote{Note that the
		experiments described in Section~\ref{evaluation} use both
		images and disks.}.
		All the ext4 structures are stored in the
		image (the superblock, the inode vector, the data blocks, etc.).
		In the configuration stage (Stage 1), this image is formatted and
		the empty logs are created.
		Thus, the log files are stored within this image, but these are
		not the \emph{real logs}. We will refer to the log files that
	 	are stored in the image as the \emph{img logs}.
		The user can create, modify, and delete any other file in this
		image. Only the pre-configured log files are WORM append-only
		files.

		\item The files provided by SealFS for the \emph{real logs}.
		These files are finally stored in a different, private block device of
		the Socarrat device, which is independent of the image file.
		SealFS authenticates the data appended to the \emph{real logs},
		enabling the tamper-evident properties by following
		a HMAC ratchet/storage-based hybrid scheme~\cite{sealfsv1,sealfsv2}.

		Finally, there are two kind of files in this private volume:
		(i) The \emph{real log files}
		and (ii) the authenticated log file (also known as
		$SEAL_{log}$, see~\cite{sealfsv1}) that contains the
		metadata of all the append operations performed
		over the \texttt{log} files.
		In the audit stage (Stage 3),
		the \emph{real logs} can be verified with $SEAL_{log}$.

		\texttt{Soca} does not export any operation
		to read or modify (delete or overwrite) the committed data
		of the \emph{real logs}, or access the $SEAL_{log}$ in any way.
	\end{itemize}

When \texttt{soca} detects an append operation to a log, it updates both the
\emph{real log} and the \emph{img log}. When it detects a read operation from
a log, \texttt{soca} responds with the data stored in the \emph{img log}. This way,
soca acts as a data diode for the \emph{real logs}.

To redirect the blocks to \texttt{soca}, we use NBD~\cite{nbd}.
While there are some alternatives available for Linux\footnote{We
also use the 9P protocol
for an alternative experimental implementation for the Plan~9 operating
system~\cite{9soca}.}, like
ublk~\cite{ublk}, we decided to use NBD for various reasons.
First, NDB has been available for a very long time and has been tested
thoroughly. Second, the protocol between the NBD client (which talks
to the Linux kernel, the Socarrat device's kernel in our case)
and the NBD server (\texttt{soca} in our case) makes debugging simpler.
NBD is composed of two different interfaces:

\begin{itemize}
	\item  The network interface, with a protocol
	to export block devices.  It is a client-server
	protocol.

	\item The operating system interface. It permits
	the client program (a user-space program named
	\texttt{ndb-client}) to import
	the network NBD device as a regular block device.
\end{itemize}

\begin{figure}[th]
\begin{center}
	\includegraphics[width=\columnwidth]{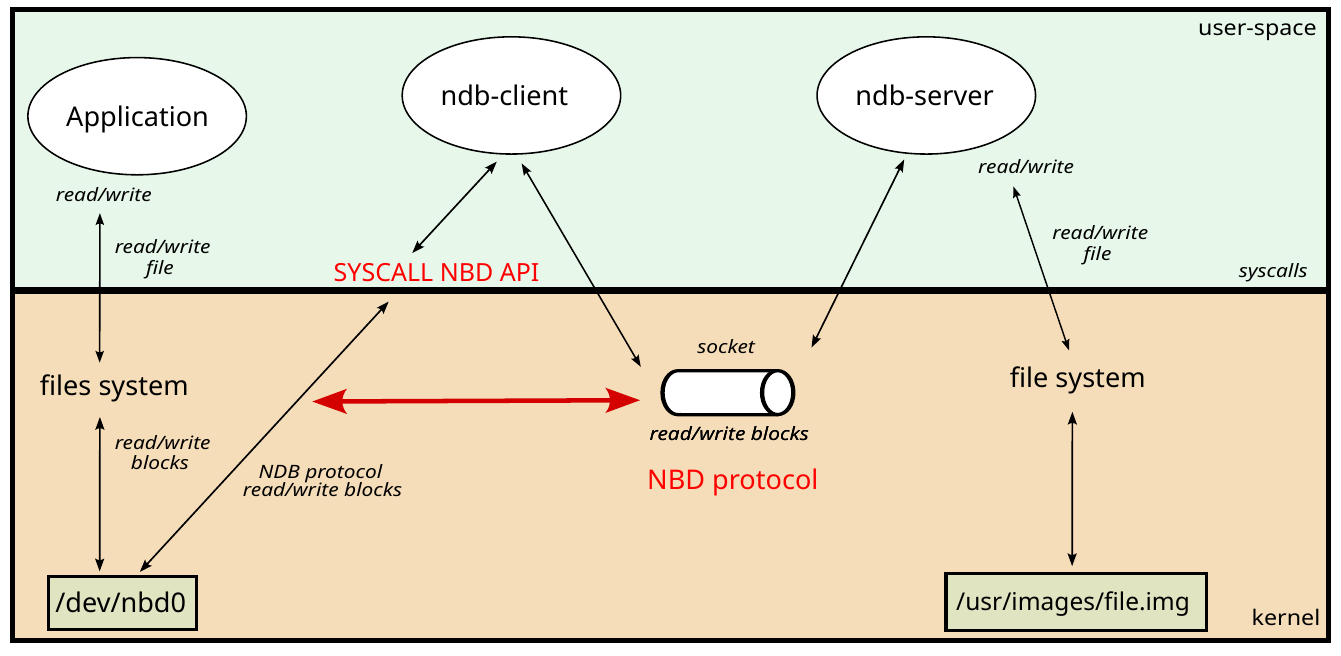}

	\caption{NBD running in a single machine. The client and the server
	use a Unix domain socket to execute the NDB protocol. \label{nbd2}}
\end{center}
\end{figure}

The client and the server can be located on the same machine, using a
Unix domain socket (UDS) to communicate, as illustrated in Figure~\ref{nbd2}.
This is the case of our architecture.

The NBD protocol
has two distinct phases: configuration and data transmission.
After the configuration phase, the client passes the socket to the
kernel and waits for the session to finish
(as depicted by the red arrow in Figure~\ref{nbd2}).
The transmission phase is performed by the kernel, that handles
the operations performed on the device that represents the device
block (\texttt{/dev/ndb0} in the figure).
In our system, this device is exported by using USB On
The Go (OTG).  To do this,
we configure a \emph{gadget}\footnote{A \emph{gadget} is
essentially the profile of the device (the identifiers, kind of endpoints,
etc.).} with \texttt{configfs}~\cite{configfs}.

Our current prototype has two modes of functioning:

\begin{itemize}
	\item \emph{Remote mode}: \texttt{Soca} acts as the
	NDB server, simply serving the Socarrat device. In this case,
	any external NDB client
	program (there are many versions available)
	can be used to connect \texttt{soca} to the Linux kernel.
	In this mode, \texttt{soca} is totally agnostic of the operating
	system and protocol it uses to connect to the NDB client.

	\begin{figure}[th]
	\begin{center}

		\includegraphics[width=\columnwidth]{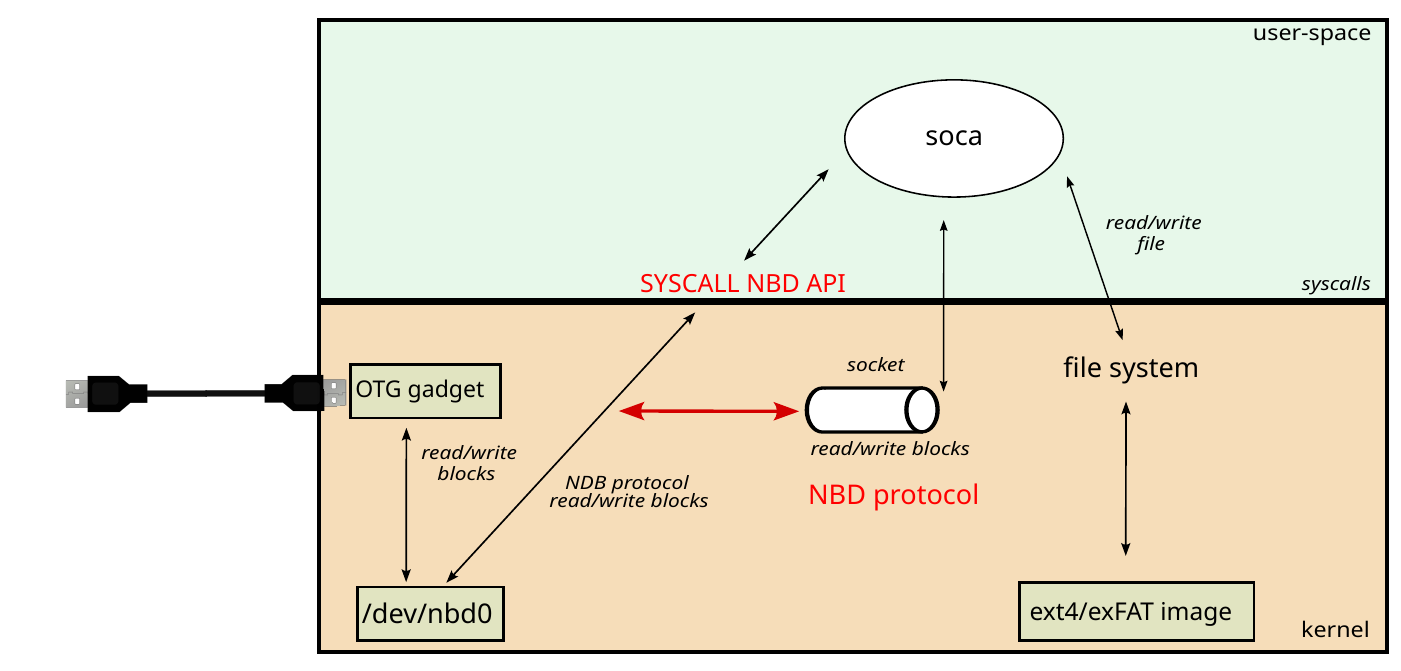}
		\caption{\texttt{Soca} in \emph{local mode}. \label{nbd3}}
	\end{center}
	\end{figure}

	\item \emph{Local mode}: \texttt{Soca}
	acts as both the NDB client and server. Figure~\ref{nbd3} depicts this mode.
	It uses the system call NDB API to
	interact with the kernel, skipping the NBD protocol
	negotiation phase for performance. Then,
	it starts the transmission phase with the kernel
	directly. This is the standard mode.
\end{itemize}

\section{Compromise detection and response\label{response}}

There are different indicators of compromise of the host machine
that can be detected by \texttt{soca}:

\begin{itemize}
	\item Receiving a write operation for a log file that is
	not a data append. In this case, the host machine is violating
	the \emph{append-only} attribute of the file system

	\item Receiving modifications of the log file metadata that are not compatible
		with append-only files (e.g. decrementing the size of the file).

	\item Detecting modifications of some file system structures (e.g. some
		parts of the superblock).

	\item Volume exhaustion. Note that we assume that the storage devices
	of the system are big enough to operate during Stage 2.
	If the file system within the image runs out
	of space, we can consider that the system is compromised.
\end{itemize}

When \texttt{soca} detects any indicator of compromise, it can follow
two different approaches:

\begin{itemize}
	\item \emph{Read-Only Remount}: Forcing the file system to remount in a
	read-only mode prevents the attacker from writing to any file on
	the Socarrat device, although they may be aware of the detection.

	\item \emph{Honeypot Mode}: Activating a honeypot mode seals the \emph{real logs},
	making them immutable. The attacker is able to modify any file on the image
	(including the \emph{img logs}) without noticing, causing the \emph{real logs}
	and \emph{img logs} to diverge. Consequently, the attacker forges what they
	believe are the legitimate logs, remaining unaware of the detection.

	Later, during the audit phase, the differences between the \emph{real logs} and
	the \emph{img logs} can be analyzed to identify the data the attacker attempted
	to modify or delete from the logs.
\end{itemize}

\section{Reverse File System: Implementation \label{implementation}}

To program a \emph{Reverse File System}, we need access to the following resources:

\begin{itemize}
	\item The block device where the file system being
	reversed is located (i.e. the image).

	\item The stream of block read/write operations.

\end{itemize}

Normally, this would be the job of a block device driver (e.g. a kernel-space component).
For fast prototyping and ease of debugging, we wanted
to be able to run this driver in user-space.  Our current \texttt{soca}
prototype is a user-space program written in Go. It has around $8394$
lines of code at the time of writing this paper.

Socarrat caches heavily. It is a concurrent program written carefully to
not block any operation unless it is strictly necessary.
It is programmed following the CSP
model~\cite{csp} (i.e. communicating threads via channels),
sharing state by communicating. It also uses \emph{mutexes} for some shared state.

As stated before, the prototype understands two
popular file systems, ext4~\cite{ext4} and exFAT~\cite{exFAT}.  Ext4 is
the native Linux standard file system. ExFAT is widely used standard file
system for removable devices and it is supported on all major operating
systems (Windows, Linux, and macOS). Supporting both provides a good
balance between portability (exFAT) and being able to experiment with
advance features (ext4).
Note that, in order to develop the \emph{Reverse File System},
we had to implement libraries to access, serve, and modify
the file systems (i.e. we have implemented simplified user-space versions of the
ext4 and exFAT drivers).

\subsection{Ext4}

Ext4 is based on \emph{inodes} (like the traditional Unix file system).
Simplifying, the partition in which the file system resides is composed of
two parts: (i) the metadata describing the file system (superblock
and group descriptors) and (ii) the data of the file system itself (the
inode vector, which is composed of inodes and an allocation bitmap, and
the data blocks and its allocation bitmap). Inodes represent file system
objects, directories and files, and contain all the metadata except
for the name. Data blocks contain data for files and directory
entries for directories. A directory entry contains the name and the
inode number\footnote{It also contains the inode type, but this is
an optimization and the information is also on the inode itself.}.

An inode contains pointers to a subtree whose leaves are
the data blocks.
There are different ways to configure how this subtree behaves.
\texttt{Soca} configures the ext4 file system to use extents,
which are simply contiguous regions of blocks.
In this configuration, the tree is a tree of extents.
This approach increases the performance by enabling the reading
of contiguous blocks from the disk.

\texttt{Soca} watches the read/write operations on
the blocks of the \emph{img log}'s  inode. When it grows, the new data
added to the last blocks referenced by the inode is appended to the
\emph{real log} (which is not served to
the user's machine).

The prototype supports ext4 with and
without journaling.
A file system with journaling has a special data structure on disk (the
journal) where it records updates to the file system transactionally, before
committing them to disk. In case of an unexpected reboot (for example from
a loss of power) the file system can transition to a consistent state
by completing or undoing a partial update.  The file system can provide
different guarantees depending on what is recorded in the journal.
If the journal only records metadata, the file system will be guaranteed
to be coherent (the tree will not contain broken links, loops, orphan
subtrees, etc), but some of the file's data may be lost or made up in
the case of a loss of power.  If the journal records data and metadata,
the data contained in a file may be lost but not only the file system
will be coherent, but spurious data will not appear in the file system.

\subsubsection{Coherency}

The main problem  is that the
blocks written to the file system
go through the cache and may be appear out of order.
The implementation of any \emph{Reverse File System} needs to answer the
question: How does one guarantee that the current file system state (on the block
device) provides a consistent
view of the file (\emph{img log}) to be able to update the \emph{real log}?
The answer depends on the guarantees the file system itself provides.
We have followed two approaches:

\begin{itemize}
	\item The first approach
	is to not assume anything about the file system except that,
	after a small wait subsequent to the first update of the metadata (i.e. the size), the
	data being written will be in its final position in the disk.
	To make sure the update is completed, we also wait for the file system to be quiescent for
	a small interval.
	The rationale is that a well designed
	file system server should, at some point, write the data and metadata to its final position
	and then not move it. The time interval during which a well designed and implemented
	file system stays incoherent should not be too wide.
	Otherwise, the file system will end up inconsistent very often.

	Some special programs (like defragmenters or
	file system checkers) can rearrange data later. Usually, these programs
	run at special times (i.e. not during Stage 2). Just in case, as an extra precaution,
	because we control the initial image, we initialize it to a known value (zero),
	to detect and delay any incoherency due to a partial update.

	We have tested this approach and it works well without exceptions.
	Nevertheless, there are no strong guarantees.
	Even with a reasonably wide coherency window,
	something may happen that breaks our assumptions.
	In that case, the log would be updated with zeroes and data would be lost.
	We have not seen this case in our experiments, but it is possible.

	\item To provide better guarantees, we can use the journal.
	For the purpose of writing a \emph{Reverse File System}, an important
	property of a journaling file system is that it clearly defines at what point
	in time the
	data and/or the metadata is known to be coherent on disk.

	The journal provides a
	transactional view of the updates of the file system. What parts of the file system
	are journaled (i.e. metadata or data) and when are they guaranteed to be written
	to the disk, depends on the journaling mode, as we will see later. The idea is to
	pick an instant when the journal guarantees that the updates to the \emph{img log} are coherent
	and then update the \emph{real log}.
\end{itemize}

\subsubsection{Ext4 without journaling}

When ext4 is configured without journaling, \texttt{soca} waits for the following events:

\begin{enumerate}
	\item A change of the file size resulting from a write
	to the block pertaining the inode for an \emph{img log}.

	\item A quiescent state of the file system for a short time window ($\lambda$).

	\item  The end of a small fixed time window ($\omega$).
\end{enumerate}

Both $\lambda$ and $\omega$ are configurable. We will refer to the sum of both
as $\tau=\lambda+\omega$.

After $\tau$, we update the \emph{real log}, traversing only the
subtree needed to access the correct offsets,
which should have been updated on disk.
$\tau$ must:

\begin{itemize}
	\item Be large enough to provide coherency.

	\item Be as small as possible, because it affects the time when it
	is possible to modify the Socarrat device's  memory\footnote{Note that this
	does not happen in the host, but on the other side of the USB link.}
	to delete uncommitted entries for the logs that may content
	indicators of compromise (i.e.
	traces of exploitation or privilege elevation, etc.).

	In other words, it is the maximum time an attacker has
	to exploit the host, compromise the USB link, and execute
	privileged malicious code in the Socarrat device to remove
	the evidence.
	After this time, once the \emph{real log} is written, it is
	tamper-evident.
\end{itemize}

Note that the value of $\tau$ does not affect performance. Applications
running in the host
do not have to wait at all, because the data is already on the Socarrat device
and the block operation request (i.e. the USB operation) has already
been processed.

\subsubsection{Ext4 with journaling}

Ext4 has support for journaling, in a format called jbd2~\cite{jbd2},
which permits various levels of transactional guarantees.
There are three possible modes of journaling for ext4 other than not
having journaling:

\begin{itemize}
	\item \emph{journal\_data}: Data and metadata are committed to the
	journal prior to being written to the file system.

	\item \emph{journal\_data\_ordered}: Only metadata is committed to
	the journal.  Data is flushed to disk before the
	data is committed to the journal.

	\item \emph{journal\_data\_writeback}: Only metadata is committed to
	the journal. Metadata is journaled but there is no guarantee
	with respect to data.
\end{itemize}

Additionally, ext4 has support for journaling some special operations
which work at a higher level of abstraction  and are kept on a
different part of the journal (i.e. \emph{fast commits}). We disabled this feature
to simplify the journal access.
\texttt{Soca} supports the following modes:

\begin{itemize}
	\item \emph{journal\_data\_ordered}.
	This mode would be preferred, because it theoretically
	provides all the guarantees we need. In this case, at the point of time when
	the metadata of a \emph{img log}
	is completely updated, data should already written to the disk. Therefore,
	\texttt{soca} can update the corresponding \emph{real log}.
	Unfortunately, empirically, if the data is being updated continuously,
	it may persist on cache and this guarantee is violated.
	From the documentation, it is a challenge to ascertain if this is a
	implementation error or this guarantee is weak.

	For this reason, in this mode,
	\texttt{soca}
	takes advantage of the journal, but at the point where it would
	update the \emph{real log}, it follows an approach similar to the
	one described for the non journaling case (i.e. it waits for the
	file system to be quiescent, etc.).
	This works well with the ext4 driver implementation and
	provides weak guarantees (those of ext4).

	\item \emph{journal\_data}. In this case,
	\texttt{soca} provides strong guarantees.
	It just watches the journal and updates the \emph{real log}
	whenever changes
	are completely committed to the journal for the corresponding
	\emph{img log}.
	Note that, in this case, \texttt{soca} does not wait $\tau$.
\end{itemize}

The prototype processes operations in strict order,
so that dependencies in the journal data
blocks do not cause an order inversion.

\subsection{ExFAT}

ExFAT is a hybrid between a FAT (i.e. a table-based linked list allocated file system)
and a contiguous allocation file system. The partition is roughly comprised
of the boot sector with the metadata of the file system, the FAT regions
containing the FAT table (the list of file system
blocks, which they call \emph{clusters}),
and the blocks region (called cluster heap).
Directory entries contain all the metadata of the file/directory
and are in the data clusters of the directories. Directory entries
for files point to their data clusters. There are essentially (again,
simplifying) two type of directory entries: contiguously allocated and
FAT allocated. Whenever posible, as an optimization, files are contiguously
allocated.

ExFAT does not provide journaling. It
supports an extension called TexFAT which
adds transaction-safe operational semantics, but it is only implemented
on some operating systems (like Windows CE). \texttt{Soca} does not support
it.

Thus, \texttt{soca} simply acts like in the case of ext4 when
there is no journal.
Whenever
the directory entry for a \emph{img log} is modified and the file size changes,
it waits $\tau$ and updates the \emph{real log}.
To read the data written to the \emph{img log}, it traverses the
prefix of the subtree (which may mean accessing the disk contiguously or using the FAT table
depending on the contents of the directory entry).

\section{Evaluation\label{evaluation}}

\subsection{Testing}

We have taken a four level testing strategy:

\begin{itemize}
	\item Go unit tests for each file system type. Some tests use
	our user-space implementation of the drivers
	used by the \emph{Reverse File System}.
	Other tests exercise the
	file systems by using a block device which injects a zero block
	every other time. This way, we check that
	the implementation correctly detects uninitialized  blocks.

	\item Go integration tests. Those tests run the whole system as an
	integrated NBD client an server (\emph{local mode}), attacking
	it concurrently. The objective of these tests is to stress the
	implementation.

	\item Shell integration tests. They use the standard Linux
	NBD client and the kernel to serve the block device, mounting it
	locally, with the kernel acting as a client. This way, we check
	that our implementation of the \emph{Reverse File Systems}
	is compatible with the kernel file system drivers.

	\item Manual integration tests through the USB link. Those
	tests check the system when in normal operation.
\end{itemize}

We tested different values for $\tau$. Empirically, in our hardware, several
hundreds of milliseconds are enough to pass the tests. For regular operation,
we set a value of $1\,s$ to provide a security margin.

\subsection{Measurements}

In all the experiments,
\texttt{Soca} has been configured in \emph{local mode}.
It sets $\lambda=10\,ms$ and $\omega=1\,s$.

To evaluate our prototype, we conducted experiments using
a slightly modified version of the standard Filebench
benchmark~\cite{filebench}. Filebench was modified to work with
append-only files, following the same approach used to evaluate
SealFS~\cite{sealfsv1,sealfsv2}. The modification has a negligible
performance impact.
Note that Filebench is run for $60\,s$ on each test.

In all scenarios, the host is a 12th Gen Intel(R) Core(TM)
i7-1280P with 20 cores and 48 GB of RAM, running Ubuntu Linux 24.04,
with a Linux kernel 6.8.0.

We tested Socarrat on four different architectures:

\begin{itemize}
	\item Raspberry Pi 4 model B running in 64-bit mode.  It is
	based on a BCM2711, with a quad core Cortex-A72 (ARM v8)
	64-bit SoC with a clock running at 1.8GHz and 4GB of RAM.
	It has a USB 2.0 OTG
	port (high-speed).  Its operating system is Debian 12 (Bookworm)
	with a Linux kernel 6.12.34.  The storage is an SD card SanDisk
	Ultra 10 HCI, with read speeds of 98 MB/s and write speed of 10
	MB/s (Class 10 rating with Ultra High Speed Class 1 rating).
	This is the cheapest device evaluated in the experiments
	($\approx 50$ \euro{}).

	\item Raxda ROCK 4c$+$ in 64-bit mode. It is based on RK3399-T,
	which has 6 cores: a dual Cortex-72 at 1.5GHz and a quad
	Cortex-A53 at 1.0GHz. It has 4GB of RAM and a USB
	3.0 OTG port (super-speed). The storage is an SD card SanDisk
	Extreme, microSDXC Class 10 U3 A2 V30 with 170 MB/s of read
	speed and 90 MB/s of write speed.  Its operating system is an
	Armbian 25.5.0-trunk.550 (based on debian Bookworm)
	with a Linux kernel 6.12.28.

	\item Raxda ROCK 5b in 64-bit mode. It is based on RK3588,
	which has a quad Cortex-A76 at 2.2~2.4GHz and a quad Cortex-A55
	at 1.8GHz. It has 12 GB of RAM a USB 3.0 OTG port
	(super-speed). The storage is an SD card SanDisk Extreme,
	microSDXC Class 10 U3 A2 V30 with 170 MB/s of read speed and 90
	MB/s of write speed.  Its operating system is an Armbian
	25.5.2 (based on debian Bookworm) with a Linux kernel 6.1.115.

	\item Orange Pi 5 Ultra in 64-bit mode. It is also based on RK3588,
	with the same processors as the Raxda ROCK 5b,
	and 16 GB of RAM. It has a USB 3.0 OTG port
	(super-speed). The boot storage is a SD card SanDisk Extreme PRO, Class V30
	U3 A2 XCI, with 200 MB/s of read speed and 120 MB/s of write speed.
	The additional storage is a 1 TB WD BLACK SN850X NVME drive.
	Its operating system is an Armbian
	25.5.2 with a Linux kernel 6.1.115.
	This is the most expensive device evaluated in the experiments
	($\approx 230$ \euro{}).
\end{itemize}

We had to backport SealFS to the kernel 6.1 to make it work in the
ROCK 5b and the Orange Pi 5 Ultra (it worked out of the box for the
other devices).  Even though there are other versions of Armbian with
more modern kernels, USB OTG is not supported yet.

The Raxda ROCK 5b architecture has been measured with two configurations:

\begin{itemize}
	\item Both the Socarrat image and the \emph{real logs}
	stored on the boot SD Card (labeled \texttt{raxda5b} in the plots).

	\item The Socarrat image stored on an
	external USB SSD disk
	and the \emph{real logs} stored on the boot SD Card
	(labeled \texttt{raxda5b\_disk} in the plots).
\end{itemize}

The Orange Pi 5 Ultra architecture
has been measured with two configurations:

\begin{itemize}
	\item The Socarrat image stored on the NVME disk and the
	\emph{real logs} stored on the boot SD Card
	(labeled \texttt{orangepi5nv} in the plots).

	\item Both the Socarrat image and the \emph{real logs} stored
	on the NVME disk (labeled \texttt{orangepi5\_allnv} in the plots).
\end{itemize}

In the rest of the architectures, both the Socarrat image and the
\emph{real logs} are stored on the boot SD Card.

\begin{figure}[p]
\begin{subfigure}{0.45\textwidth}
  \centering
  \includegraphics[width=\linewidth]{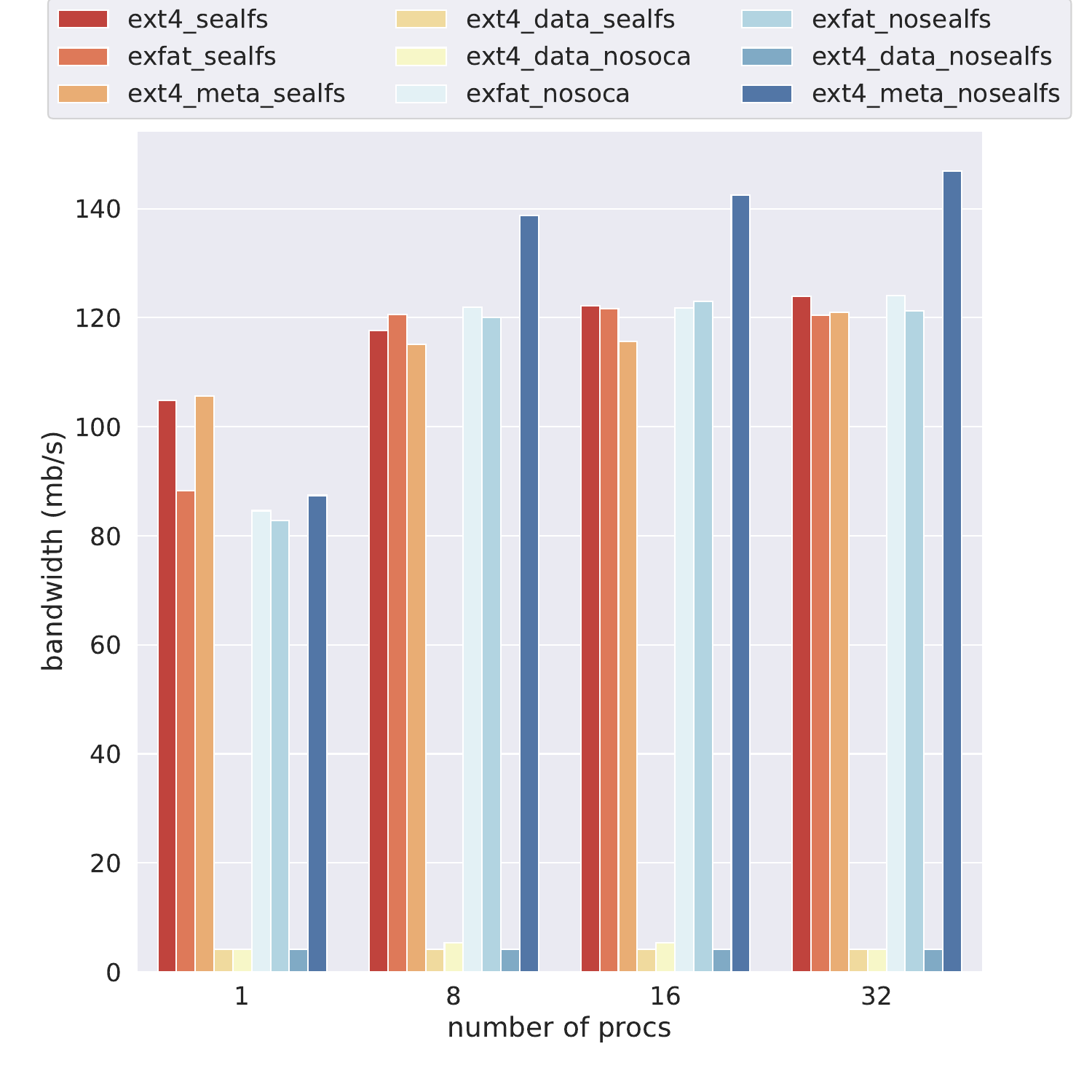}
  \caption{Raspberry Pi 4}
\end{subfigure}%
\begin{subfigure}{0.45\textwidth}
  \includegraphics[width=\linewidth, viewport=0 0 709 620, clip]{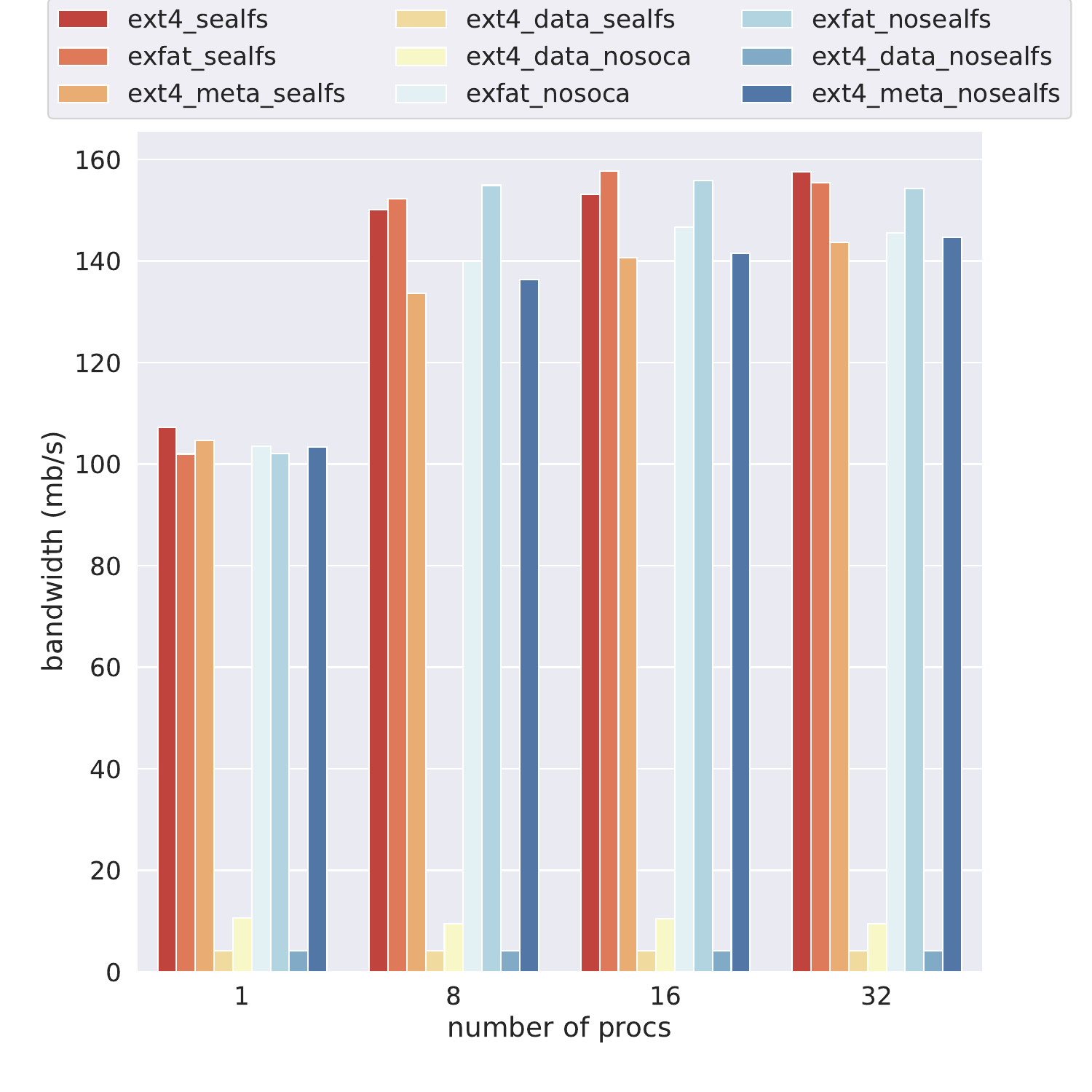}
  \caption{Raxda 4c Plus}
\end{subfigure}\\
\begin{subfigure}{0.45\textwidth}
  \centering
  \includegraphics[width=\linewidth, viewport=0 0 709 620, clip]{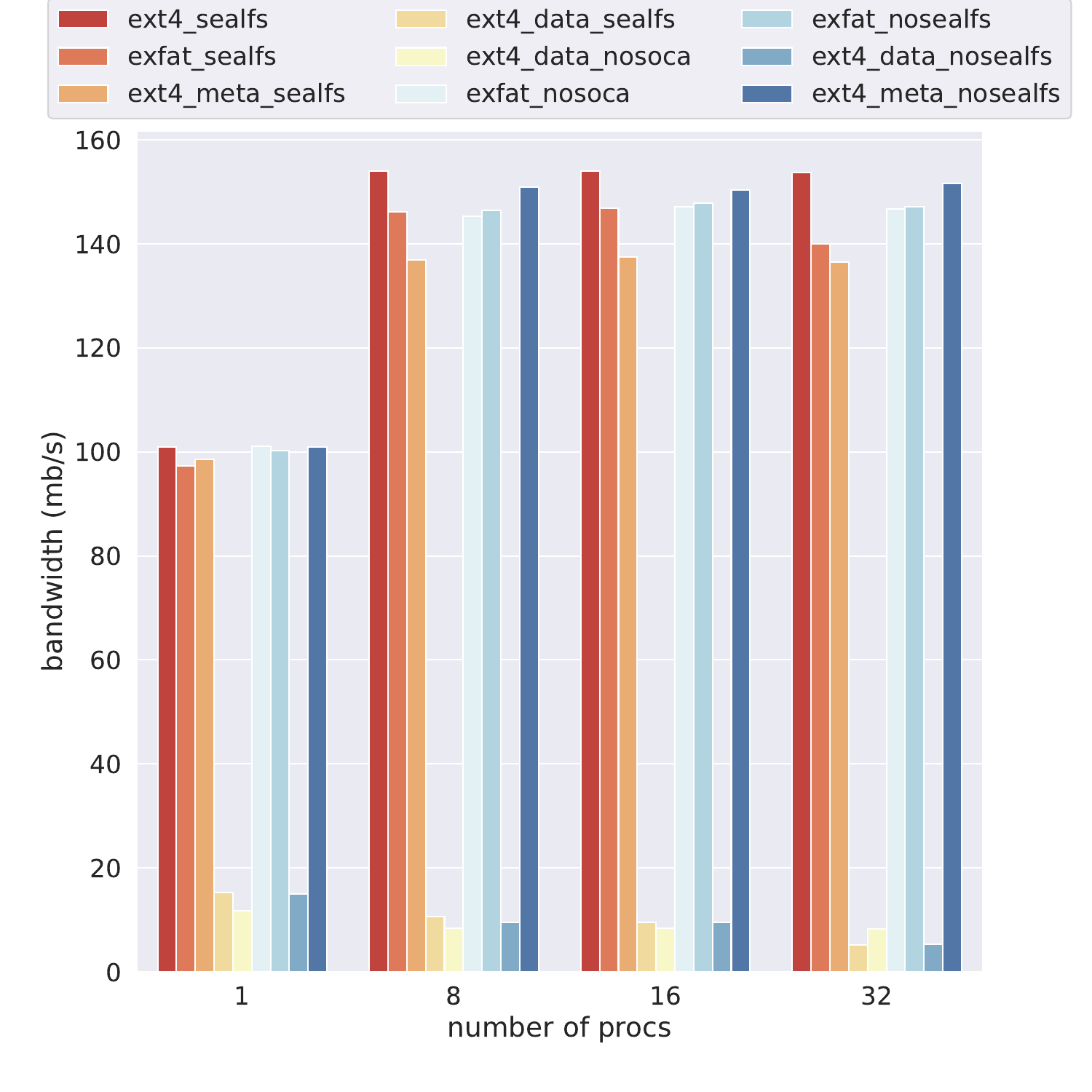}
  \caption{Raxda 5b (SD Card)}
\end{subfigure}%
\begin{subfigure}{0.45\textwidth}
  \includegraphics[width=\linewidth, viewport=0 0 709 620, clip]{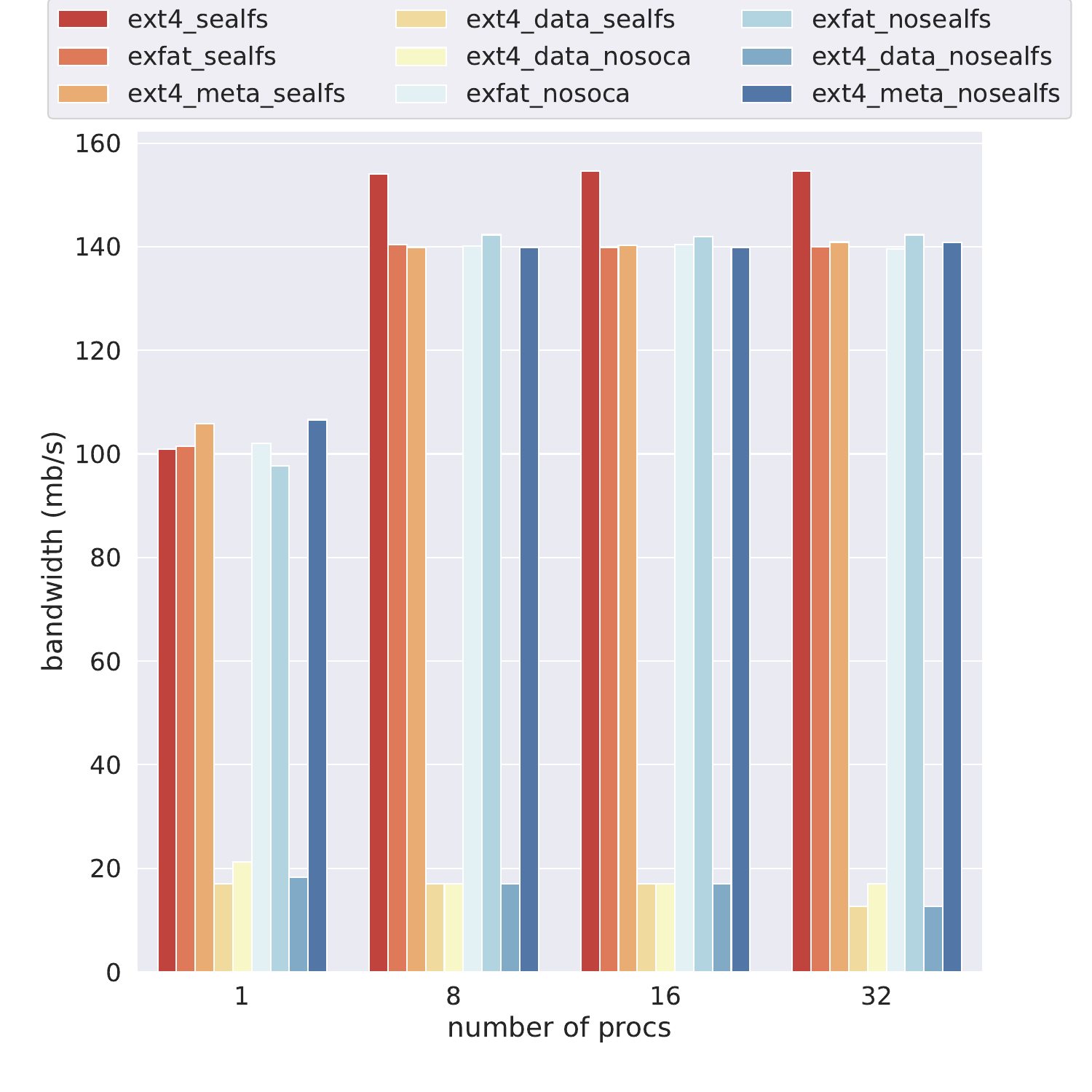}
  \caption{Raxda 5b (SD Card and USB SSD)}
\end{subfigure}\\
\begin{subfigure}{0.45\textwidth}
  \centering
  \includegraphics[width=\linewidth, viewport=0 0 709 620, clip]{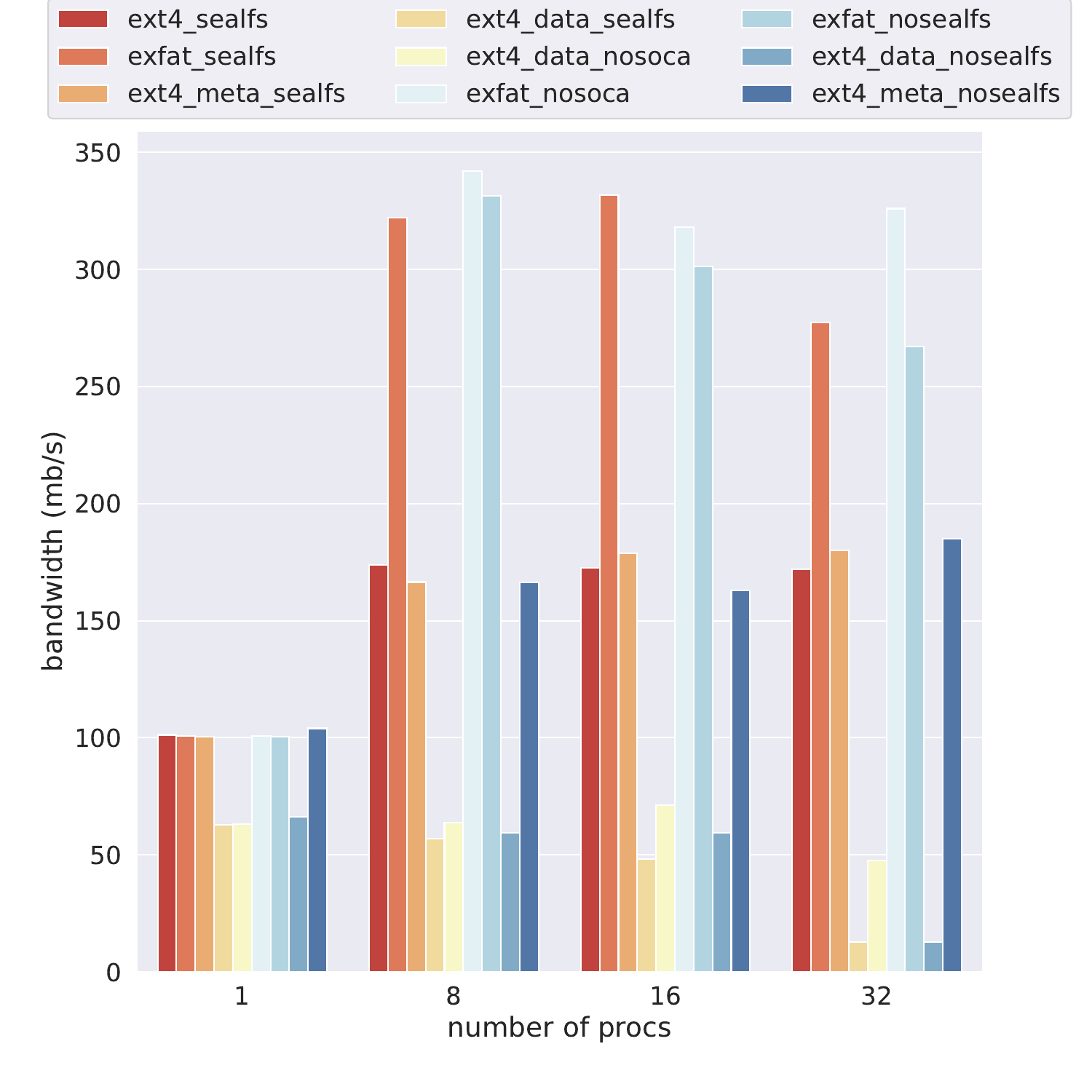}
  \caption{Orange Pi 5 Ultra (NVME)}
\end{subfigure}%
\begin{subfigure}{0.45\textwidth}
  \includegraphics[width=\linewidth, viewport=0 0 709 620, clip]{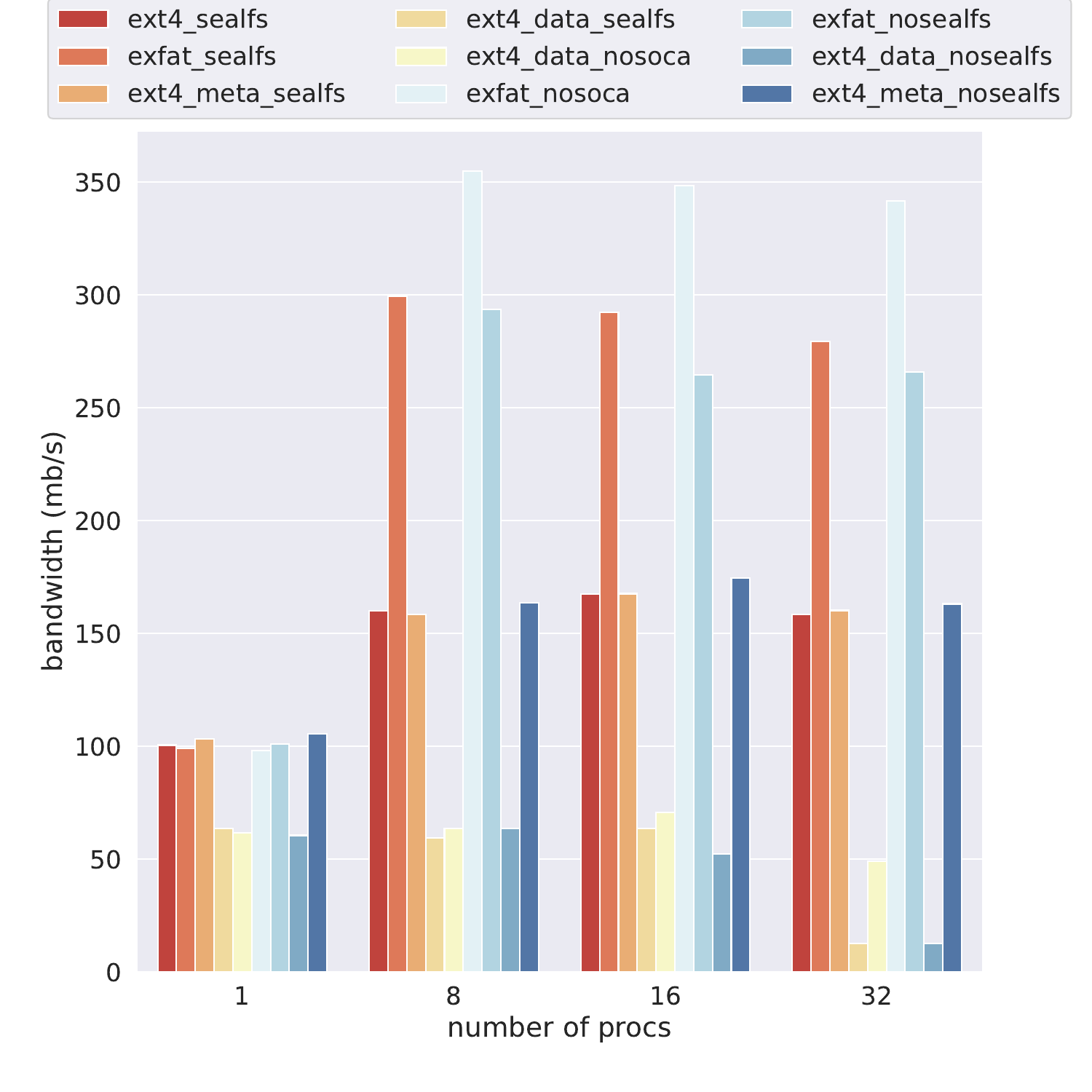}
  \caption{Orange Pi 5 Ultra (NVME and SD Card) \label{bw:orange}}
\end{subfigure}\\
\caption{Bandwidth measured by Filebench. \label{bw}}
\end{figure}

\begin{figure}[p]
\begin{subfigure}{0.45\textwidth}
  \centering
  \includegraphics[width=\linewidth]{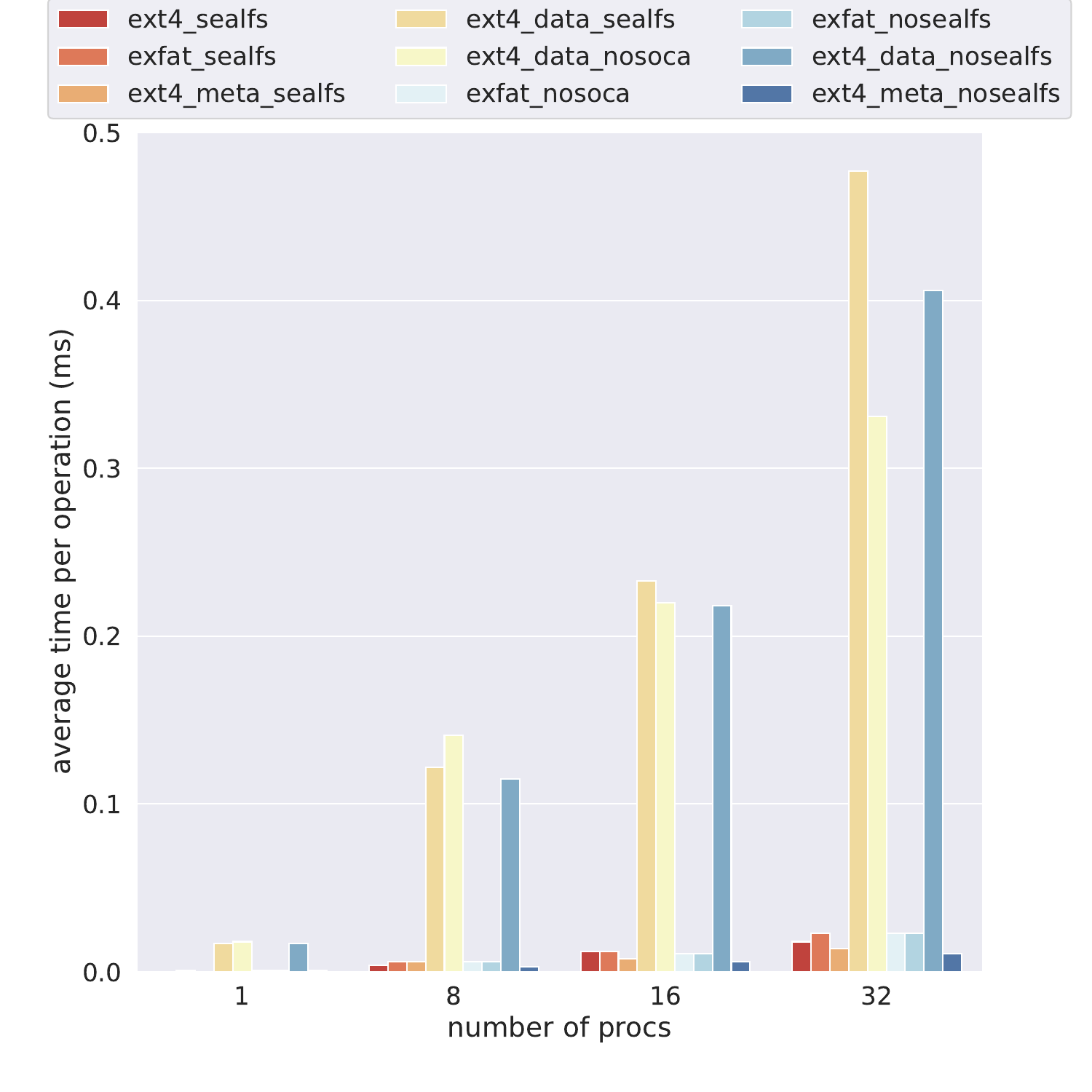}
  \caption{Raspberry Pi 4}
\end{subfigure}%
\begin{subfigure}{0.45\textwidth}
  \includegraphics[width=\linewidth, viewport=0 0 709 640, clip]{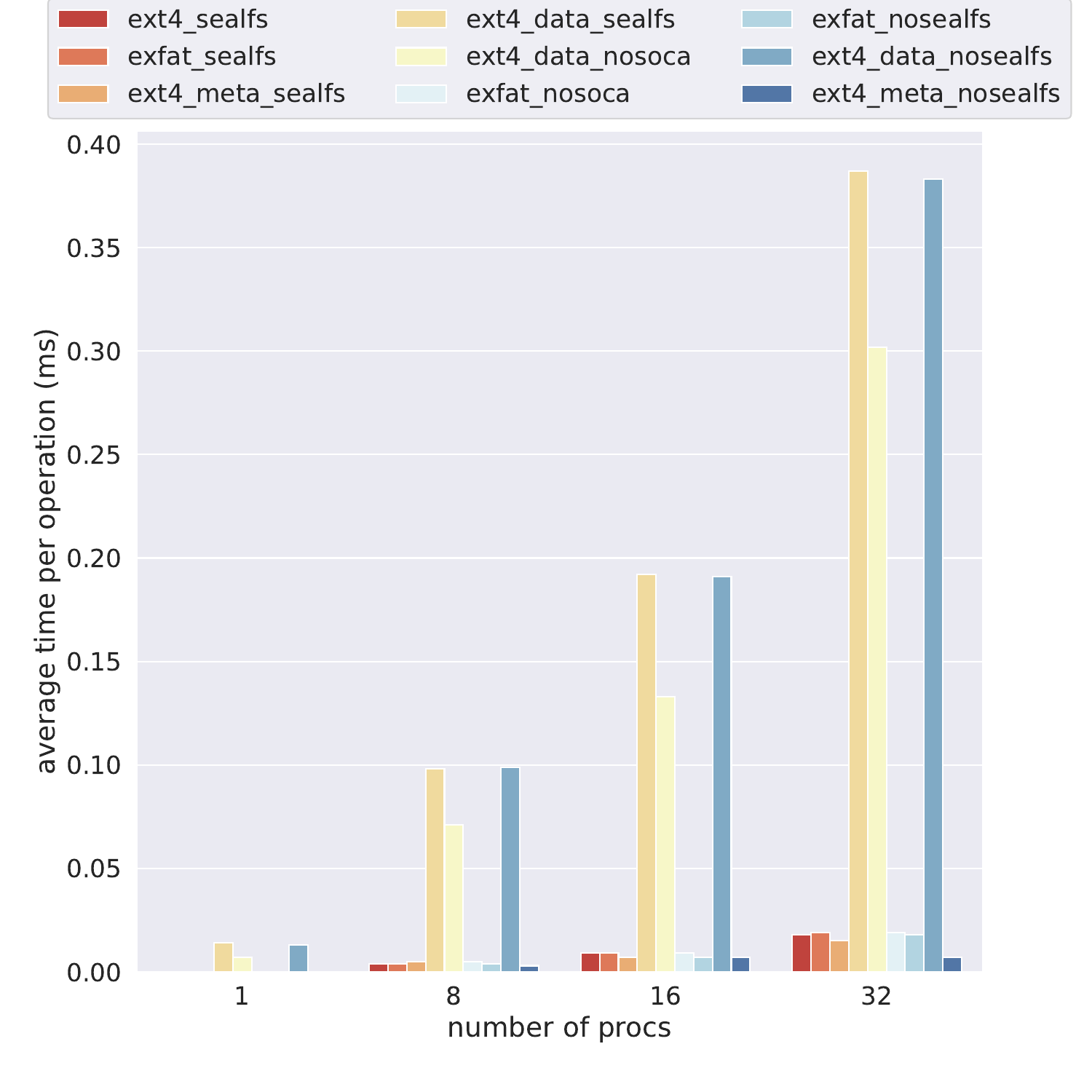}
  \caption{Raxda 4c Plus}
\end{subfigure}\\
\begin{subfigure}{0.45\textwidth}
  \centering
  \includegraphics[width=\linewidth, viewport=0 0 709 620, clip]{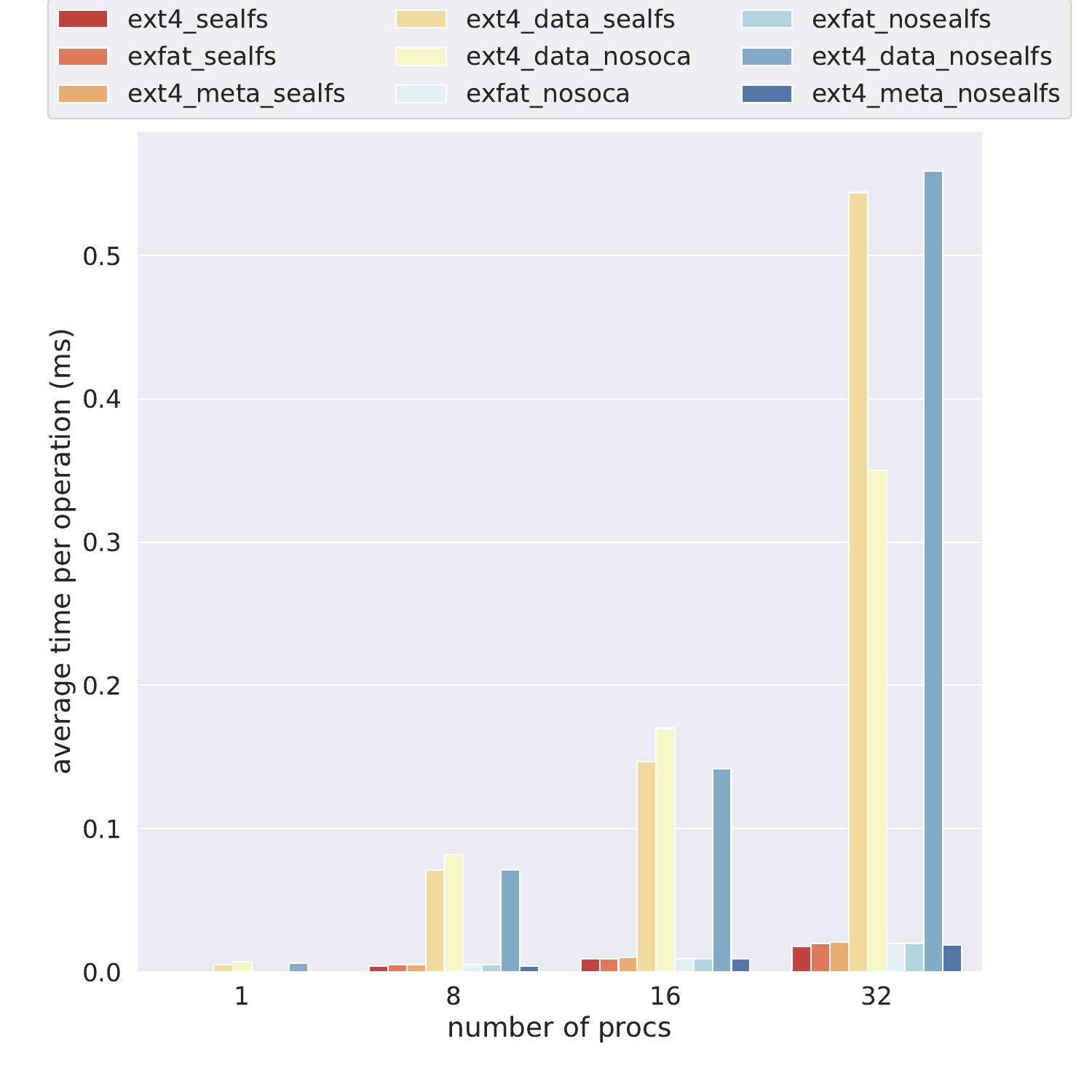}
  \caption{Raxda 5b (SD Card)}
\end{subfigure}%
\begin{subfigure}{0.45\textwidth}
  \includegraphics[width=\linewidth, viewport=0 0 709 620, clip]{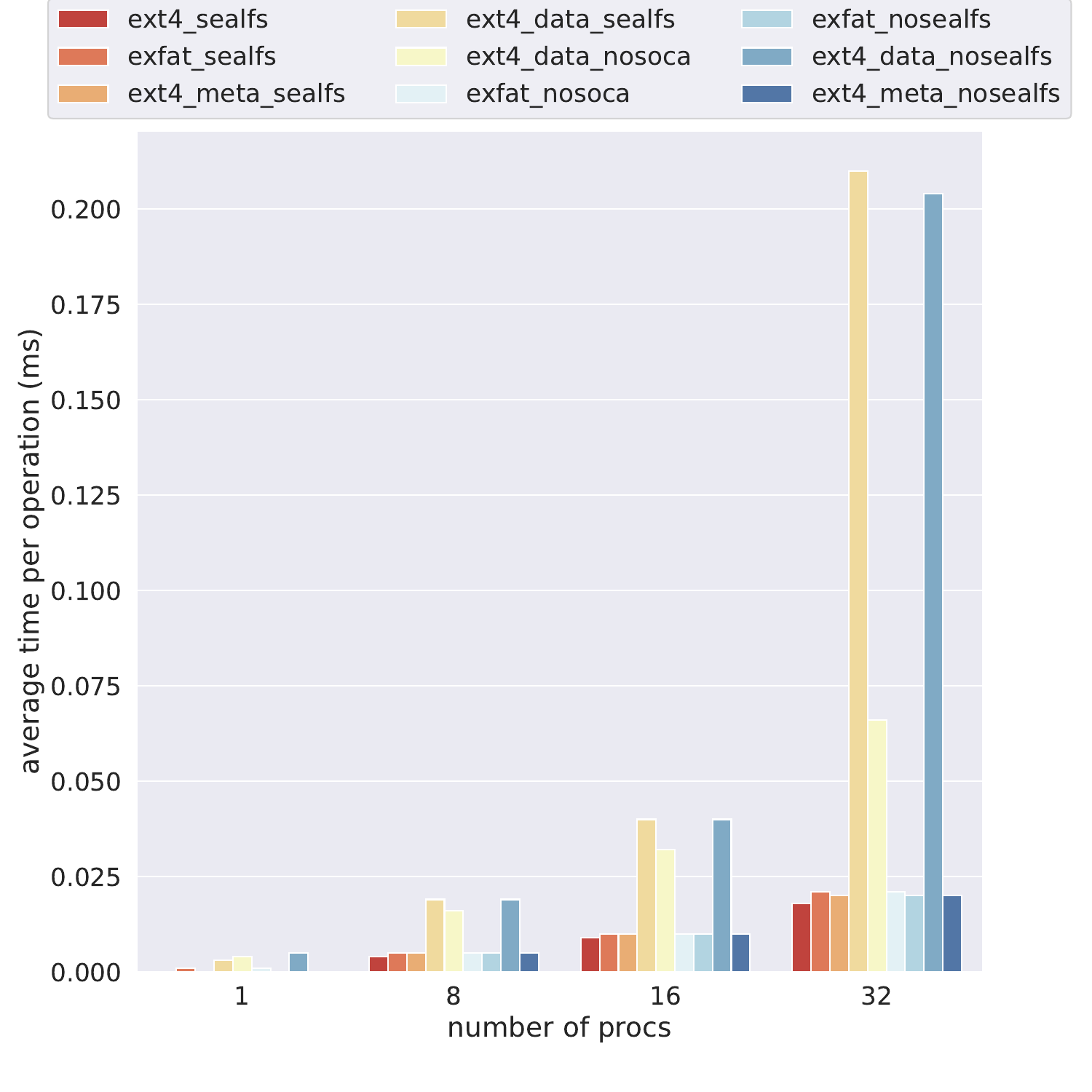}
  \caption{Raxda 5b (SD Card and USB SSD)}
\end{subfigure}\\
\begin{subfigure}{0.45\textwidth}
  \centering
  \includegraphics[width=\linewidth, viewport=0 0 709 620, clip]{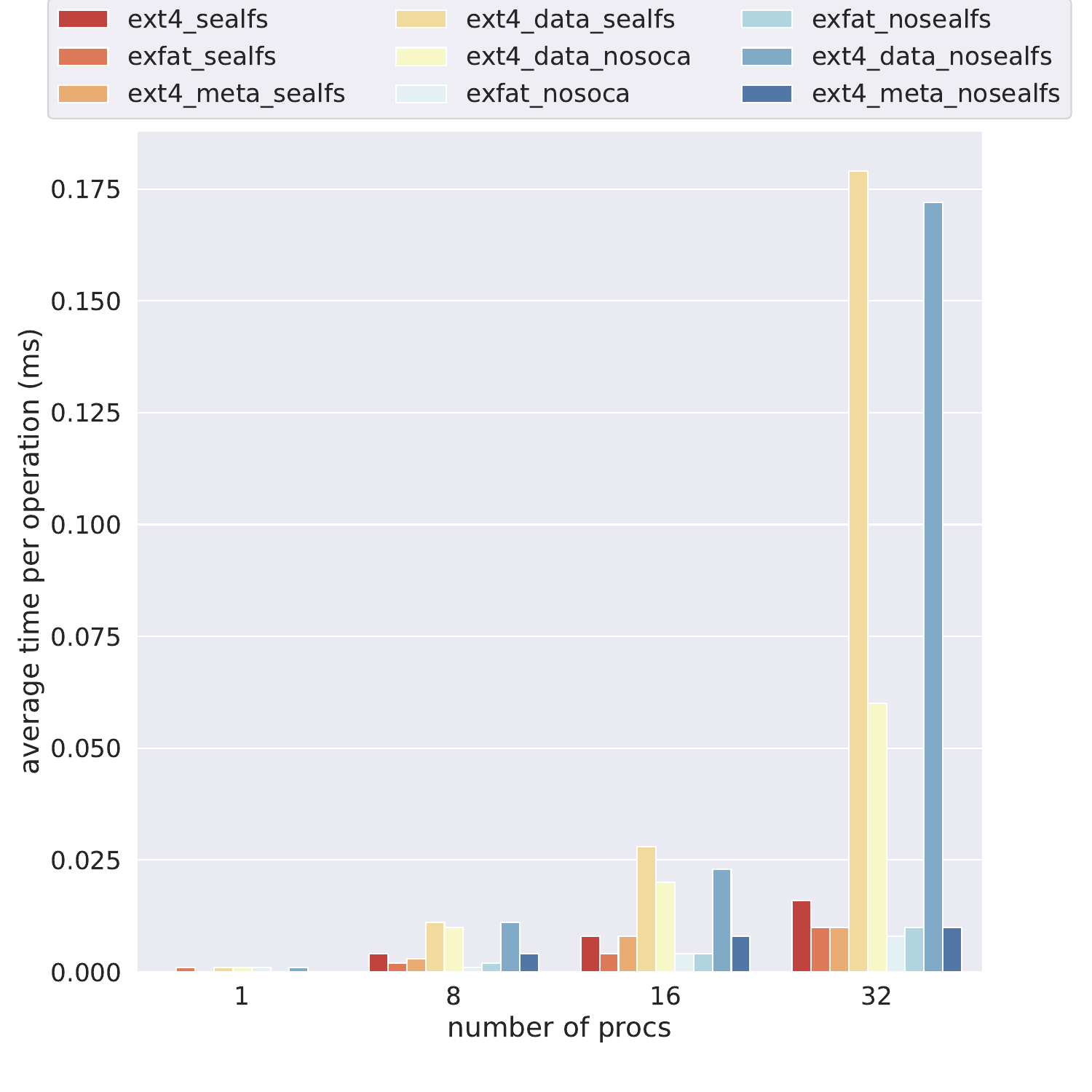}
  \caption{Orange Pi 5 Ultra (NVME)}
\end{subfigure}%
\begin{subfigure}{0.45\textwidth}
  \includegraphics[width=\linewidth, viewport=0 0 709 620, clip]{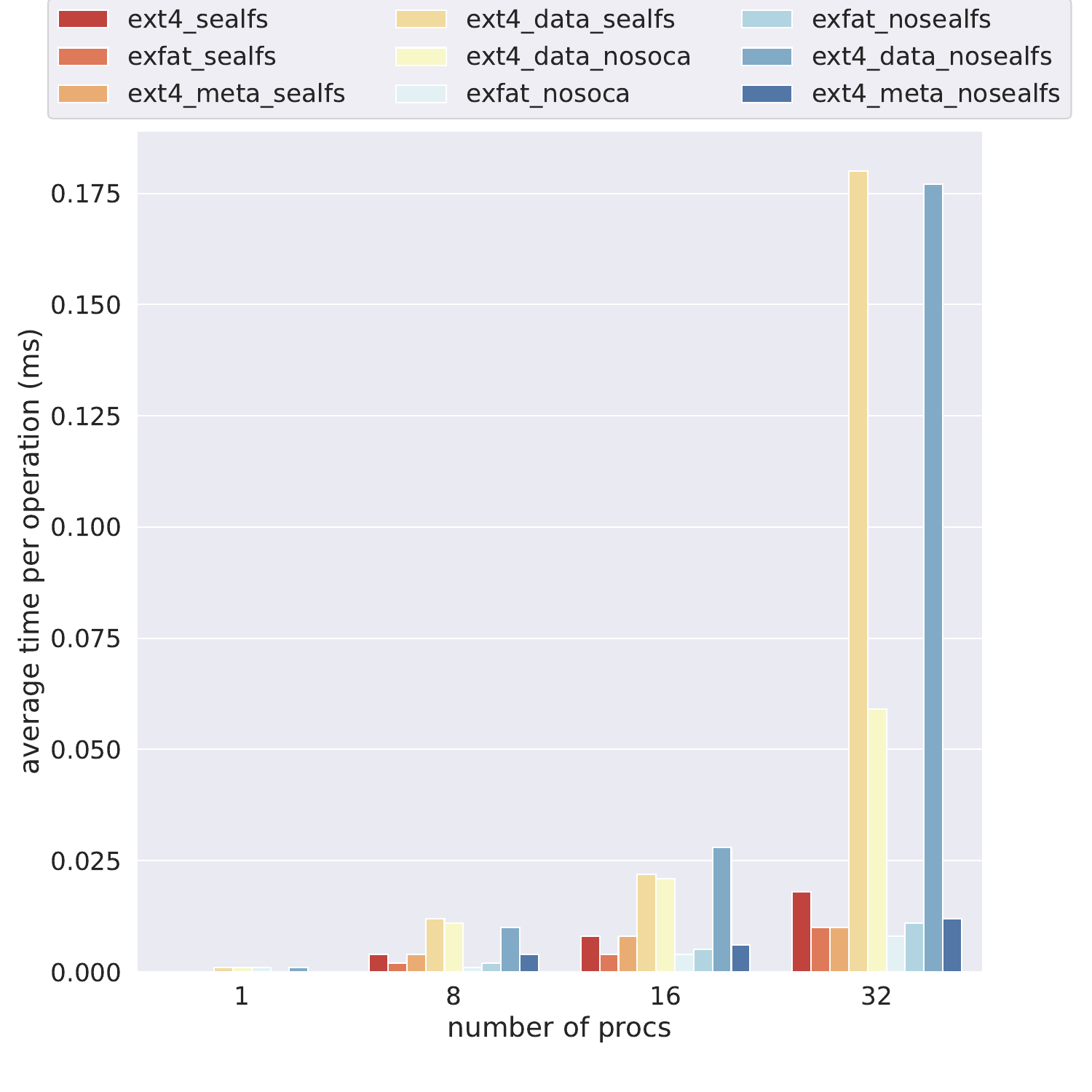}
  \caption{Orange Pi 5 Ultra (NVME and SD Card) \label{lat:orange}}
\end{subfigure}\\
\caption{Average latency measured by Filebench. \label{lat}}
\end{figure}

Figure~\ref{bw} and Figure~\ref{lat} show the results of the measurements
performed by Filebench for different number of concurrent processes.
Figure~\ref{bw} shows the bandwidth and Figure~\ref{lat} shows
the latency (the average time for an operation).
In those plots:

\begin{itemize}
	\item Bars labeled \texttt{\_nosoca} depict measurements
	where the device exports directly an image filesystem file using USB OTG.
	That is, this is the baseline for the measurements, because \texttt{soca}
	is not involved (i.e. there is no \emph{Reverse File System}).
	Note that, in this case, the block device is exported
	directly (there is  no NBD).

	The other cases export an NBD device served by soca using USB OTG.

	\item Bars
	labeled \texttt{\_nosealfs} use soca but do not keep the \emph{real logs}
	under a SealFS mounted directory. They provide a baseline to measure
	the impact of SealFS. Bars labeled with \texttt{\_sealfs}
	use soca with SealFS.

	\item Bars labeled \texttt{ext4\_meta} use
	ext4  with a journal configured as
	\emph{journal\_data\_ordered}.

 	\item Bars labeled \texttt{ext4\_data} use
	ext4 with a journal configured as \emph{journal\_data}.

	\item The rest of bars labeled \texttt{ext4\_} use ext4
	without journaling.

 	\item Bars labeled \texttt{exfat\_} use exFAT.
\end{itemize}

\begin{figure}[tp]
 \includegraphics[width=\linewidth]{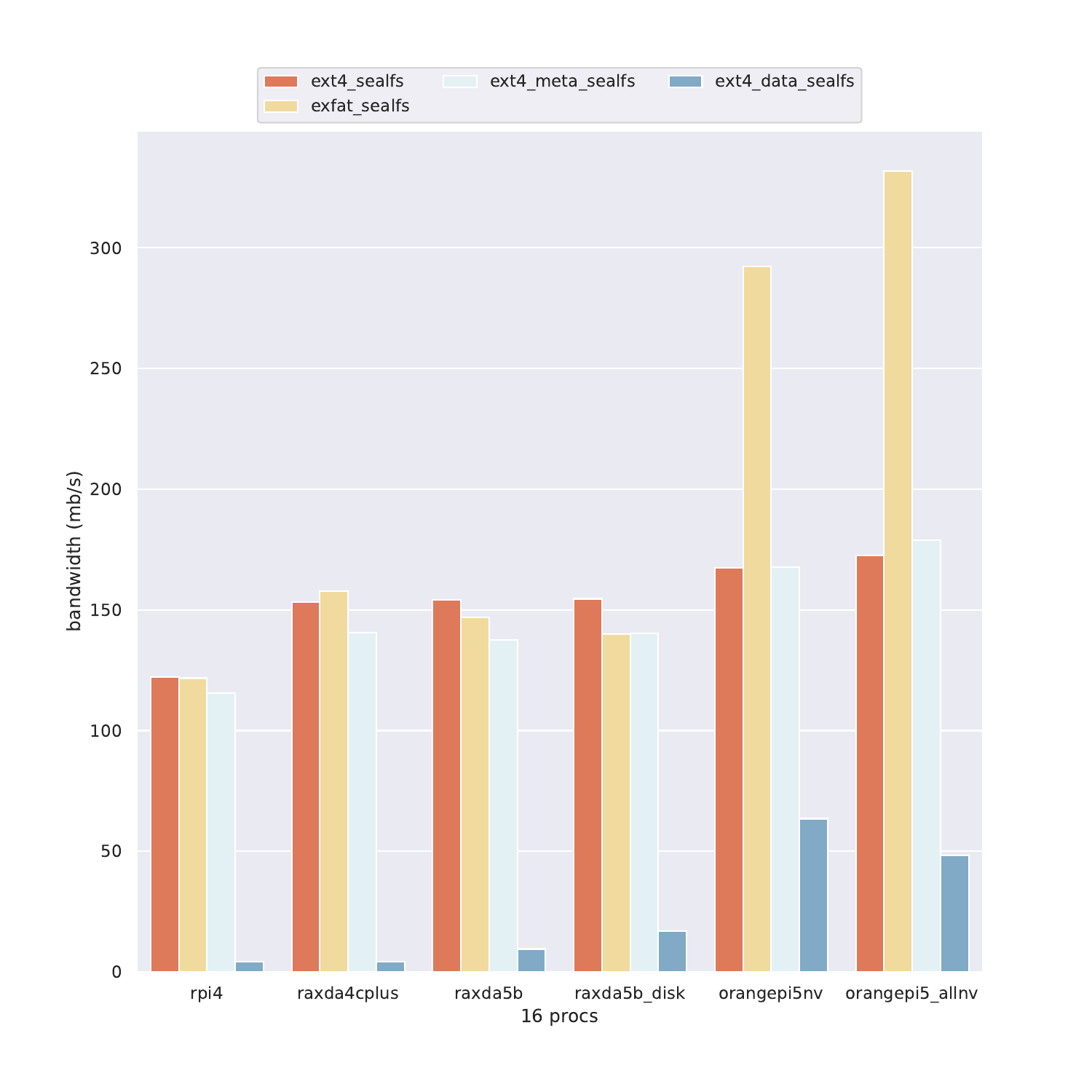}
\caption{Comparison of the results of Filebench (bandwidth) for 16
concurrent processes in different platform configurations.}
\label{fig:beta}
\end{figure}

\begin{figure}[tp]
 \includegraphics[width=\linewidth]{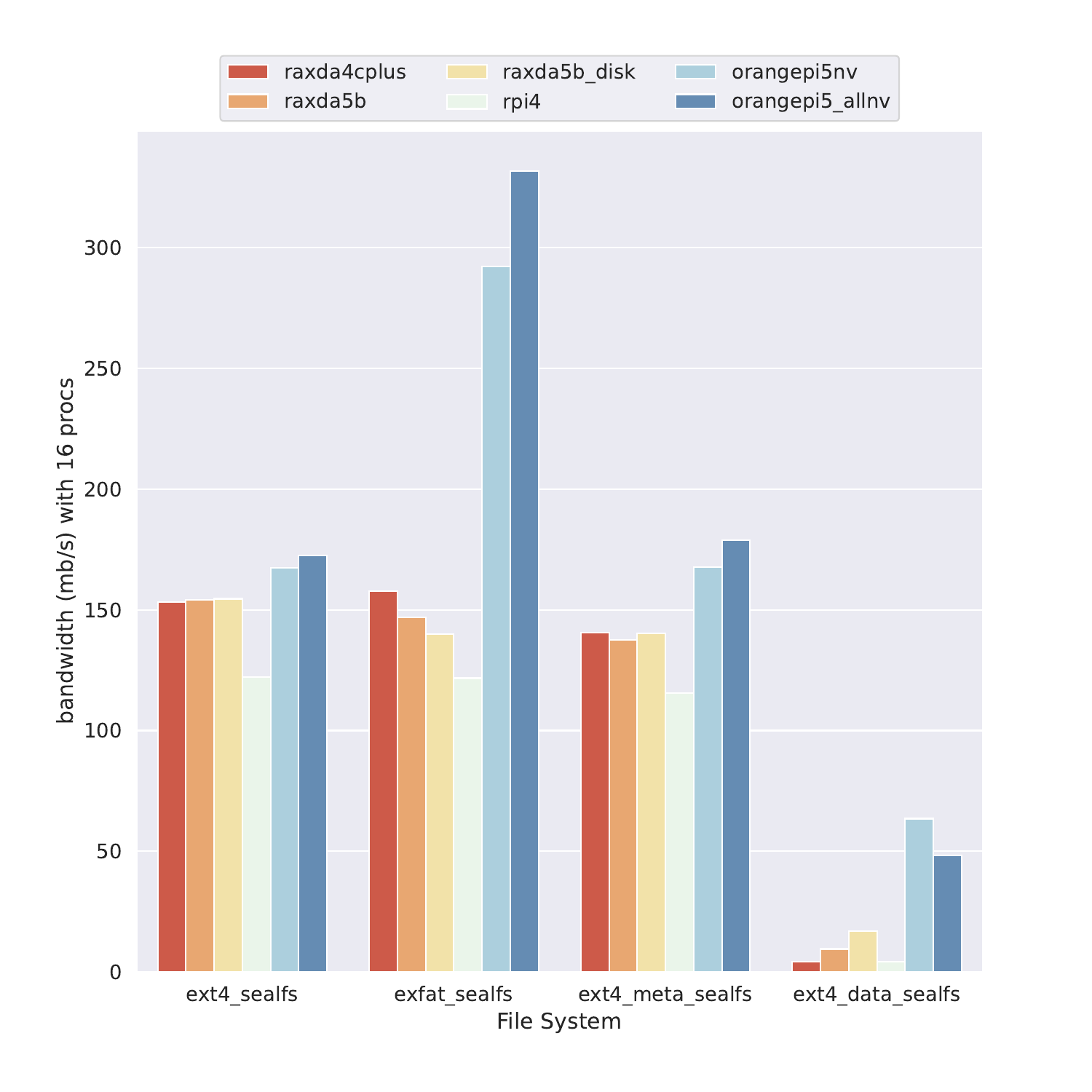}
\caption{Comparison of the results of Filebench (bandwidth) for 16
concurrent processes, for different file systems.}
\label{fig:gamma}
\end{figure}

Figures~\ref{bw} and Figure~\ref{lat} reflect the impact of concurrency
in the host.
As the host has 20 CPU cores, we have selected a constant number of
processes (16 processes, less than 20 cores) for the next figures,
which are focused on bandwidth.
Figure~\ref{fig:beta} shows the bandwidth results
for 16 concurrent processes, grouped by hardware configuration.
Figure~\ref{fig:gamma} shows the bandwidth results,
grouped by file system configuration.
Those figures only depict the file system configurations with
all the security properties explained in our model (i.e. Socarrat and
SealFS).

\subsection{Analysis and discussion}

Note that this is a complex system of two interconnected machines with
many different hardware and software components involved, so analyzing
the results is not trivial.

Regarding the latency, it grows with the number of processes in all
cases. This is due to contention in the host device (e.g. spin locks
in the kernel). When the number of processes is greater than the number
of CPU cores, the performance is notably worse. In the case of ext4 with
data journaling, the latency is extremely worse. This is due to the same
effect that we will describe later, in the bandwidth analysis.

The bandwidth is worse for
one process (versus multiple processes), in all cases. One plausible
explanation is that the host is underusing the Socarrat device. While
the host is waiting for the response of a USB request, the link is
idle, because there are not other concurrent requests to process.
With more processes injecting requests, there is bigger utilization
of the capabilities of the device (so the bandwidth grows).

In general, the bandwidth is not worse with 32 processes, even when the host has 20
CPU cores. This is expected, because they are I/O bound processes.
Nevertheless, in some cases there is exhaustion with 32 processes. This is
the case of ext4 with data journaling in the most powerful architectures.
In this cases, the journal is the bottleneck.

There is a clear performance gap between \texttt{ext4\_data} and the
rest of file systems. This is the cost of strong guarantees.
If we want the strong guarantees offered by the journal, \texttt{soca}
has to wait synchronously for the points in time where the storage device
commits the data and the file system is coherent.
Moreover, the journal is a circular data structure that gets full and we
have to wait for parts of it to be evicted before overwriting them, which
compounds this effect (it is much worse when there is data is also journaled).

When only the metadata is journaled, there
is not so much traffic.
Note that in the worst cases (Raspberry Pi 4 and Raxda 4c$+$),
the performance is greater or equal than $4.2\,
MB/s$, which should be more than enough for a regular logging scenario.
In the best scenario, we provide
more than $63.5\,MB/s$. Note that recommended bandwidth to
stream Ultra High Definition (UHD) video is 15 Mbps or
higher\footnote{See for example Netflix recommendations:
\texttt{https://help.netflix.com/en/node/306}}.

As expected, the performance of ext4 with journaling
improves in the cases with faster storage devices (e.g. NVME disk) and
dual storage configurations (i.e. mixing the SD Card and the NVME disk).
In the later case, performance benefits may be derived from the load
balancing effect of having two different drives.

A surprising result is that, in some cases, using Socarrat
is slightly faster than exporting the block device directly.
Note that we measured the later as a baseline.
This could be a consequence of two compounded effects:

\begin{itemize}
	\item \texttt{soca} caches heavily (in different ways, including keeping
	a queue of answered but not attended operations).

	\item A regular block device in Linux has a small block size
	(normally between 4 KB and 16 KB).
	NBD devices, on the other hand, can handle bigger writes
	of contiguous data, making operations more efficient.
\end{itemize}

The effect of these factors along the chain ends up delivering
bigger chunks to write to disk at the other end, which is also
more efficient.

Comparing architectures (Figure~\ref{fig:beta}),
the best hardware in all cases is the Orange Pi 5 Ultra, as expected.
Surprisingly, the Raspberry Pi 4 (the cheapest and more widely available
configuration) is not much worse than the Raxda 4c$+$, even though it only
supports USB OTG 2.0.

Regarding file systems, the fastest is exFAT, because there is no journaling
and the blocks (i.e. clusters) are bigger, and allocated contiguously when
possible. Also, the file system data structures are simpler (i.e. it requires
updating less blocks per write).

SealFS does not affect latency or bandwidth significantly, as can be observed
in Figure~\ref{bw} and Figure~\ref{lat}.

\section{Conclusions\label{conclusions}}

In this work, we have presented Socarrat, a novel solution for implementing
Write Once Read Many (WORM) storage devices using a low-cost, widely available
single-board computers. Socarrat achieves strong data immutability guarantees
without relying on proprietary hardware or complex distributed infrastructures.

Our approach, based on the \emph{Reverse File System}, enables transparent
integration with standard file systems such as ext4 and exFAT, requiring no
modifications to the host's operating system or additional drivers.
Socarrat not only ensures WORM append-only files, but also incorporates
tamper-evident features that
enhance its resilience against physical compromise.

The system is fully implemented in Go and released as free/libre software,
promoting transparency and reproducibility.
Our evaluation demonstrates that Socarrat exhibits reasonable performance
when running on cost-effective hardware: It can be practically employed for secure logging
purposes, even in demanding scenarios such as video logging.

As future work, it would be interesting to implement the local Linux
interface using ublk~\cite{ublk} instead of NBD. It would probably have
performance benefits (less copies of memory and system call
batching, due to the io\_uring interface). Another possible change would
be to switch the local Linux interface to netlink communication with
the kernel~\cite{netlink}.
Another idea left as future work is the possibility of sealing files.
Once a file is done with, it could be sealed and become immutable.
For example, in ext4,
this can be done by changing its attributes.
Once a file is sealed, a trail is written to it (so that SealFS logs it)
and \texttt{soca} closes it. This file is considered
read only and the real log file for it cannot be changed.

Socarrat is free/libre software (GPL).
The source code is available in the our
public repositories:

\begin{center}

 \texttt{\url{https://gitlab.eif.urjc.es/publications/soca}}

\end{center}

\section*{Competing Interests}

There are no competing interests in this work.

\section*{Funding}

This work is funded under the Proyectos de Generación de
Conocimiento 2021 call of Ministry of Science and Innovation of
Spain co-funded by the European Union, project
PID2021-126592OB-C22 CASCAR/DMARCE.

\section*{Acknowledgments}

The authors would like to thank Fermín Galán Márquez
for constructive criticism of the manuscript.

Generative AI software tools (Microsoft
Copilot\footnote{https://www.bing.com/chat}) have been used exclusively
to edit and improve the quality of human-generated existing text.


\end{document}